\documentclass{aastex61}
\usepackage{natbib}

\usepackage{graphics,graphicx,amsmath,amssymb,amsthm,epsfig}
\def\kms{km~s$^{-1}$}
\def\OMC{$O-C$}

\received{2016 November 8}
\accepted{2017 April 3}
\submitjournal{ApJ}

\shorttitle{RED NOVA PREDICTION}
\shortauthors{Molnar et al.}

\begin{document}
\title{PREDICTION OF A RED NOVA OUTBURST IN KIC 9832227}

\correspondingauthor{Lawrence Molnar}
\email{lmolnar@calvin.edu}

\author{Lawrence A. Molnar}
\affiliation{Department of Physics \& Astronomy \\
	Calvin College, Grand Rapids, MI 49546, USA}

\author{Daniel M. Van Noord}
\affiliation{Department of Physics \& Astronomy \\
	Calvin College, Grand Rapids, MI 49546, USA}

\author{Karen Kinemuchi}
\affiliation{Apache Point Observatory \\
	Sunspot, NM, 88349, USA}

\author{Jason P. Smolinski}
\affiliation{Department of Physics \& Astronomy \\
	Calvin College, Grand Rapids, MI 49546, USA}

\author{Cara E. Alexander}
\affiliation{Department of Physics \& Astronomy \\
	Calvin College, Grand Rapids, MI 49546, USA}

\author{Evan M. Cook}
\affiliation{Department of Physics \& Astronomy \\
	Calvin College, Grand Rapids, MI 49546, USA}

\author{Byoungchan Jang}
\affiliation{Department of Physics \& Astronomy \\
	Calvin College, Grand Rapids, MI 49546, USA}

\author{Henry A. Kobulnicky}
\affiliation{Department of Physics \& Astronomy \\
	University of Wyoming, Laramie, WY 82072, USA}

\author{Christopher J. Spedden}
\affiliation{Department of Physics \& Astronomy \\
	Calvin College, Grand Rapids, MI 49546, USA}

\author{Steven D. Steenwyk}
\affiliation{Department of Physics \& Astronomy \\
	Calvin College, Grand Rapids, MI 49546, USA}

\begin{abstract}
We present the first identification of a candidate precursor for an imminent red nova.
Our prediction is based on the example of the precursor to the red nova V1309 Sco,
which was retrospectively found to be a contact binary with an exponentially decreasing period.
We explore the use of this distinctive timing signature to identify precursors, developing the observational and analysis steps needed.
We estimate that our Galaxy has roughly 1--10 observable precursors.
Specifically, we lay out the observational case for KIC 9832227, which we identified as a tentative candidate two years ago (Molnar et al. 2015, AAS Meeting Abstracts 415.05).
Orbital timing over the past two years has followed the tentative exponential fit.
As of late 2015, the period time derivative went beyond the range found in other systems ($\dot P < |1\times 10^{-8}|$), a necessary criterion for a serious candidate.
We estimate time of merger is the year $2022.2\pm0.7$.
Double absorption line spectra confirm directly the 0.458~d light curve period is a contact binary system and yield a mass ratio $m_B/m_A = 0.228\pm0.003$.
Closer analysis of the Kepler timing data shows evidence of a component C with orbital period $P_C=590\pm 8$ days and $m_C\sin i_C = 0.11~M_\odot$.
An alternative interpretation of the long term timing trend, light travel time delay due to orbit around a distant component D, is ruled out by the spectroscopic data for any nondegenerate star.
Additional measurements are needed to test further the merging hypothesis and to utilize fully this fortuitous opportunity.

\end{abstract}


\section{INTRODUCTION}\label{secIntro}

\subsection{Transient Events in Stellar Evolution}

Stellar evolution, both for individual stars and for multiple star systems, consists of both short- and long-lived phases.
Transient events are short lived-phases that are sufficiently brief and luminous to be discovered by the time variation of their brightness.
Ideally, the study of transient events would include a full characterization of the preoutburst state.
Such data would provide observational tests of outburst mechanisms as well as a fundamental check that the correct pool of progenitors is being connected to that transient type.
Unfortunately, it has not been possible, to date, to {\it predict} the occurrence of any major transient type.
This is not surprising owing to the rarity of events and the low luminosity of precursors.
Moreover, theories do not generally predict specific observational signatures of systems in the final years before eruption.

With an ever increasing observational emphasis on surveys, 
retrospective identification of precursors has been possible in some fortunate cases, allowing limited but instructive tests of theory.
Prediscovery photometry of the Type II supernova SN 1987A in the Large Magellanic Cloud showed a blue supergiant---a surprising result that led to significant reconsideration of the Type II supernova mechanism \citep[cf.][and references therein]{1989ARA&A..27..629A}.
Prediscovery photometry of the Type IIb supernova SN 1993J in M81 showed a similarly surprising color \citep{1994AJ....107..662A}.
Prediscovery light curves from the Optical Gravitational Lensing Experiment (OGLE) survey of the classical nova V1213 Cen (Nova Centauri 2009) showed a very low quiescent level punctuated with dwarf novae outbursts in marked contrast to the brighter, steadier post-nova behavior \citep{2016Natur.537..649M}.
This observation directly supports models that suggest the classical nova outburst abruptly alters the mass transfer rate.

Luminous red novae have only recently been identified as a distinct class of stellar transient \citep{2007Natur.447..458K}.
They are characterized by relatively long outbursts, a range of luminosities intermediate between classical novae and supernovae, and red spectra.
Prediscovery OGLE light curves of the luminous red nova V1309 Sco (Nova Sco 2008) revealed the light curve shape of a contact binary system with an orbital period that was decreasing exponentially \citep[][hereafter Te11]{2011A&A...528A.114T}.
This was a remarkable confirmation of the hypothesis of \citet{2003ApJ...582L.105S} that luminous red novae arise from merging contact binary stars.

Had intensive, targeted observations of the precursors been possible in any of these cases, more specific details of the physical state of the precursor could have been determined and used to test theory more fully.
For red novae, Te11 suggested the mechanism that triggers outburst is the onset of the Darwin instability.
This occurs when the mass ratio is small enough that the companion star can no longer keep the primary star synchronously rotating via tidal interaction.
Angular momentum transferred from the binary orbit to the primary's spin changes the orbit more than the spin, leading to a runaway.
Specifically, \citet{1995ApJ...444L..41R} found this occurs at a mass ratio of 0.09 for a main sequence primary star that is mostly radiative.

The central premise of this paper is that the most valuable use of Te11's discovery may be to consider the exponentially decreasing period as the first predictive signature of an imminent nova outburst.
We consider generally how this signature should be sought in the rapidly increasing quantity of survey data that will be available in the near future.
We consider specifically the fortuitous identification of a promising candidate red nova precursor, KIC 9832227 \citep{2015AAS...22541505M}, laying out the observational evidence in hand and specifying additional observations needed to fill out the story.

\subsection{Formation and Evolution of Contact Binaries}\label{subEvolution}

\citet{2012JASS...29..145E} describes contact binaries and common envelope evolution as the two great unsolved questions of stellar evolution.
There is growing consensus on the mechanisms underlying some of the key stages of contact binary evolution.
Initial binary separations are wide (orbital periods of days to months).
Dynamic interaction with third star companions combined with tidal friction can reduce the binary period down to a day or two.
See, e.g., \citet{2007ApJ...669.1298F} for dynamical simulations and \citet{2006A&A...450..681T} for an observational survey of third star companions of close binaries.
Magnetic braking and nuclear evolution may both play roles in bringing the binary from that point into contact.
Once in contact, the stellar radii and Roche geometry do not allow for significant evolution of the orbital period.
The end result for all orbital periods is a stellar merger.
\citet{2003ApJ...582L.105S} found red novae with a wide range of lumonosities can be produced by contact binaries with a range of initial masses.

There is no consensus on the details of mass exchange between the contact binary components or on the specific mechanism that triggers the merger.
\citet{2012JASS...29..145E} favors a gradual mass transfer from the less massive (and hence smaller radius) secondary star to the primary driven by structural change of the secondary caused by the energy it receives from the primary.
The contact is sustained by magnetic braking or nuclear evolution.
This comes to an abrupt end when the mass ratio is extreme enough for the Darwin instability to trigger a merger.
Alternatively, \citet{2011A&A...531A..18S} suggests that when a binary system first reaches contact, a brief intense mass transfer event ensues that ends with the originally more massive star becoming the less massive one.
A stable contact phase only exists following the reversal.
\citet{2006ApJ...643..381D} explores dynamic mass transfer without the Darwin instability.
\citet{2016RMxAA..52..113K} explore merger triggered by a tidal runaway based on a nonequilibrium response to tidal dissipation.

\subsection{The Precursor Candidate KIC 9832227}\label{subIntroKIC}

KIC 9832227 was discovered in the Northern Sky Variability Survey (NSVS) \citep{2004AJ....127.2436W} in a catalog for RR Lyrae stars \citep{2006AJ....132.1202K}.
The nature of the variability was unclear in the NSVS data, and the light curve obtained from Kepler further complicated the 
classification.
Originally, KIC 9832227 was listed as a candidate c-type RR Lyrae star, but with a short periodicity, the target could be classified as something else \citep{2013arXiv1310.0544K}.
The Kepler light curve revealed a periodicity at 0.23 d (pulsational) or 0.46 d (orbital), but with an unusual behavior of amplitude modulation that appeared to be periodic as well.
The Kepler Eclipsing Binary Catalog \citep[and references therein]{2016AJ....151...68K} found a morphological parameter of 0.95, meaning the identification as an eclipsing binary is uncertain.

\citet[][hereafter Me15]{2015AAS...22541505M} presented measurements (color vs.\ orbital phase) that ruled out the pulsation interpretation.
Double absorption line spectra presented in the present paper confirm directly the contact binary classification.
Furthermore, Me15 found the orbital period determined from Kepler data was less than the period determined from prior ground based observations over a nine-year time baseline---a negative time derivative of the period.
The signature behavior of V1309 Sco was a negative period derivative that became more negative over time.
Timing measurements of their own from 2013 and 2014 (subsequent to the Kepler data) showed the period derivative was indeed becoming more extreme.
Me15 found all of the timing data together were well described by the exponentially decreasing period function used for V1309 Sco and identified the system as a red nova precursor candidate on that basis.

However, Me15 concluded by outlining two additional criteria that must be met before the system can be considered a likely precursor for a red nova outburst:
1) the magnitude of the period derivative should exceed the range observed for otherwise similar contact binary systems;
and 2) the possibility that the observed period changes are caused by changing light travel time as the binary orbits a distant companion must be ruled out.
In this paper we present these the photometry presented by Me15 in detail for the first time, along with follow up photometric and spectroscopic measurements made through 2016 September that address these issues.

\subsection{Assessing the Size of the Red Nova Precursor Population}\label{subPopulation}

To assess the feasibility of searching for red nova precursors, we estimate the number of precursors in our Galaxy that are currently detectable as such.
Three values must be determined: the rate of nova outbursts observed, the magnitude of the period derivative required for a precursor to stand out from other systems (the threshold period derivative, $\dot P_t$), and the length of time $\dot P_t$ is exceeded before outburst.
The expected number of precursors is simply the product of the outburst rate and the number of years the precursor exceeds $\dot P_t$.

\citet{2014MNRAS.443.1319K} estimated the Galactic rate of stellar mergers as a function of their peak luminosity.
They found that events like the V1309 Sco outburst are observed about once per decade.

Contact binaries exhibit timing variations for a variety of reasons, not all understood \citep[see][for a list]{2014AJ....147...45C}.
\citet{2006AcA....56..253K} used OGLE data to make a uniform study of 569 contact binaries with data spanning thirteen years.
They found a limit to the observed magnitude of $\dot P$ of $1\times 10^{-8}$, with typical values less than a tenth of that.
Positive and negative values occurred at similar rates, indicating the dominant period change mechanism was cyclical (such as, for example, changes caused by orbital motion around a third star).
We adopt $1\times 10^{-8}$ as our threshold $\dot P_t$.

To determine the time that the period derivative will exceed the threshold, one needs a model of the time evolution of the orbit.
\citet{1976ApJ...209..829W} and \citet{1977ApJ...211..881W} showed that an unstable phase of evolution can occur in a close binary resulting in shrinkage of the orbit and merger, but no physically realistic model of the process has yet been constructed.
We therefore use the phenomenological expression that Te11 found represented well the time dependence of the orbital period of V1309 Sco.
Specifically, they used an exponential formula of the form
\begin{equation}
P(t)=P_0\exp(\tau/[t-t_0]) ,
\label{eqnExp}
\end{equation}
where $P_0$ is the initial orbital period, $t_0$ is the time at which Equation \ref{eqnExp} goes to zero, and $\tau$ is the exponential timescale.
(Note that the period does not have to go all the way to zero to trigger an outburst---for V1309 Sco the onset of outburst preceded $t_0$ by 1.8 year.)
Hence the period derivative
\begin{equation}
\dot P(t) = -P_0\tau/(t-t_0)^2\exp(\tau/[t-t_0]) ,
\end{equation}
and $\dot P$ exceeds $\dot P_t$ from the time the threshold is reached, $t_t$, until $t_0$.
Since $\tau$ is observed to be much less than a year, the exponential factor is nearly unity until the final year and 
\begin{equation}
(t_0-t_t) \approx (P_0\tau/\dot P_t)^{1/2} .
\label{eqnTime}
\end{equation}

For V1309 Sco $(P_0,\tau)$ = (1.4456~d, 15.29~d).
Hence, when first observed by OGLE in August 2001 the period derivative, $-2\times 10^{-6}$, was already more extreme than $\dot P_t$.
Using Equation \ref{eqnExp}, the time spent in excess of threshold was 129 years.
The outburst rate times this lifetime implies there should be 13 such systems in our Galaxy currently exceeding the threshold as they approach merger.
As determined \S \ref{subsecExp} below, the precursor candidate KIC 9832227 has $(P_0,\tau)$ = (0.45796~d, 0.114~d), both lower values than for V1309 Sco, and hence $(t_0-t_t)$ of 6.2 y.
This yields an estimate of 0.6 systems exceeding threshold on average.

The typical value for $(t_0-t_t)$ probably lies somewhere in between these two cases.
The orbital period of V1309 Sco, 1.4 d, is generally considered to be unusually long for contact binaries.
Indeed W UMa systems are sometimes defined to have orbital periods less than a day.
However, catalogs of ground-based data have a sampling bias against periods near one day.
A period histogram of the 261 Kepler binary systems classified as contact binaries (Matijevic et al. morphology between 0.7 and 0.8) shows this period value to be smoothly connected with shorter values.
Fully {5\%} of the systems had longer periods.
With shorter main sequence lifetimes, the more massive stars found in longer period systems may evolve more quickly, and so may be underrepresented in the histogram relative to their birthrate.
Hence, there is no reason to think the period of V1309 Sco is distinctive of systems that merge.
The orbital period of KIC 9832227 is in the peak of the period distribution (0.25-0.50~d).

In summary, it is plausible that there are currently observable red nova precursor systems, although the number is statistically small.
Furthermore, as $(t_0-t_t)$ is only six years for KIC 9832227, it is desirable in general to identify precursors as soon as they reach $\dot P_t$ to be sure of having adequate time to study them before outburst.
Hence it is important to use optimal methods of timing analysis.

\subsection{The Plan of this Paper}\label{subPlan}

In \S \ref{secV1309-phot}, we reanalyze the V1309 Sco photometry, developing the technique we later apply to KIC 9832227.
In particular, we show that emphasizing measurement of the timing of orbital phase is more precise than direct estimates of orbital period.
In \S \ref{secKIC-phot} we present the available KIC 9832227 photometry (from ground-based archives, the Kepler spacecraft, and four years of our own ground-based observations).
We give special treatment to means of minimizing the effects of time-variable starspots on timing estimates.
We find evidence of a small third body, component C, with an orbital period of 1.7 years.
We compute an updated fit of timing data to the exponential formula.
In \S \ref{secSpec} we present spectroscopic observations from Apache Point Observatory (APO) and the Wyoming Infrared Observatory (WIRO).
In \S \ref{secAnalysis} we seek a self-consistent model of the light-curve shape and the spectra.
In \S \ref{subsecTemp} we estimate temperature and metallicity directly from the spectra and compare with temperatures estimated from Sloan colors (reported in the literature and measured by us).
In \S \ref{subsecVel} we peform a global fit to the spectroscopic data for the stellar rotational and orbital velocities for each of the two data sets.
In \S \ref{subsecBM} we use the light curves and the spectroscopic velocities to determine the physical parameters of the binary system.
In \S \ref{subsecMC} we use the spectra to place an upper limit on the mass of component C.
In \S \ref{subsecCompD} we consider and rule out an alternative model of the long term timing trend involving light travel time delays due to a massive companion, a hypothetical component D.
In \S \ref{secDiscussion} we briefly discuss three aspects of the merger process for which the case of KIC 9832227 may be instructive: the outburst trigger, the outburst timing relative to the parameter $t_0$, and changes in light curve shape prior to outburst.
Finally in \S \ref{secConclusions}, we summarize our conclusions and outline further work needed.

\section{V1309 SCO PHOTOMETRIC OBSERVATIONS}\label{secV1309-phot}

The central method of this paper is to use the preoutburst behavior of V1309 Sco as a template for identifying the next red nova in advance.
To that end this section will review what is known about the approach to nova for V1309 Sco.
First we analyze the data in the same manner as our later analysis of KIC 9832227 for direct comparison.
Second, we consider the parameters of the exponential formula used for V1309 Sco; in what ways might other systems differ?
More complete reanalysis of V1309 Sco is beyond the scope of this paper.

Te11 presented an analysis of OGLE photometry of V1309 Sco spanning 2001 August to the outburst.
They found an approximately 1.4 d periodicity with two maxima and minima typical of contact binary systems, although one of the maxima gradually disappeared in the final year.
As described in Section \ref{subPopulation}, they found the time dependence of the orbital period to be well modeled by an exponential formula (our Equation \ref{eqnExp}).
To determine orbital period they fit groups of 50 data points with the method of \citet{1996ApJ...460L.107S}.
This method expands possible light curves shapes as an N component Fourier series.
It quantifies the likelihood of the existence of periodicity after searching a multidimensional parameter space of frequency and Fourier component strengths.

In our application we have a simpler hypothesis to test: what is the orbital phase (relative to some reference ephemeris) of the segment of data given the known frequency and light curve shape?
Hence we avoid what \citet{1996ApJ...460L.107S} terms the ``bandwidth penalty'' of fitting a large number of unnecessary parameters and achieve a more precise estimate.
Data from epochs too sparse to independently fix a period can still be used to fix an orbital phase.
Additionally, fits made relative to a nominal timing model eliminate smearing of the signal due to change of the period within a segment of data.
Orbital period is found as the inverse of the time derivative of orbital phase.

We implement this approach for the OGLE data as follows.
First we remove the long term brightness trend (as did Te11), a gradual brightening of about one magnitude over five years.
We define a series of epochs (full observing seasons for the first three years when the data are sparse, and half seasons for the later years).
Second, we determine the orbital phase based on the exponential ephemeris of Equation \ref{eqnExp}.
Third, we determine the shape of a typical light curve by a Fourier fit to the data from the first half of 2004, chosen for its relatively dense observational sampling and its location near the middle of the full observation set.
Finally, we compute a least squares fit of each epoch relative to the template light curve optimizing the timing offset.
This offset, observed orbital phase minus the ephemeris phase (\OMC ), is plotted in Figure \ref{figV1309Sco-OMCresid}.
The offsets are labelled in orbital cycles on the left and equivalent time in minutes on the right.
The zero point is determined by the template used, in this case the first half of 2004.
The vertical error bars are one sigma uncertainties while the horizontal ones show the time range included in the epoch.  The data points are marked at the average time for data in each epoch.
There is a systematic delay of 0.07 cycles that occurs from 2004 to 2005.
The lack of a large change in 2006--2007 shows the method can be remarkably tolerant of changes in the light curve shape.

The same data are replotted in Figure \ref{figV1309Sco-OMC} relative to a fixed reference period, along with a line representing the exponential formula.
Data points are placed at average times for each epoch as before.
On this plot the timing uncertainties are smaller than the symbols. 
This is a more typical presentation of timing data.
A constant period appears as a straight line.
A positive (negative) slope indicates a period longer (shorter) than the reference period.
A constant negative (positive) period derivative has the shape of a concave down (up) parabola.
The reference period was chosen to be the actual period near the middle of the data set.
The transition from small curvature of the V1309 Sco data in the early years to large curvature in the later years indicates the magnitude of the period derivative is increasing.

A straight line can be fit to each pair of adjacent data points from Figure \ref{figV1309Sco-OMC} and the slope used as an estimate of the orbital period at the time midway between the points, with uncertainties determined by combining the timing uncertainties in quadrature.
These are plotted in Figure \ref{figV1309Sco-P} along with a line representing the exponential formula.
For reference, the maximum period of the plot range is $P_0$, the initial period of Equation \ref{eqnExp}.
By comparison with the period versus time plot of Te11 (their Figure 2), the uncertainties are substantially smaller, and the time range covered greater (as the 2001 data are used for our first data point).
Even with this greater precision, the agreement with the exponential formula is quite good (in the sense that deviations of the measured periods from the formula are small compared to the net change in period).
The greatest deviation from the model is the high point in 2004, which corresponds to the offset at the same year in the residual \OMC\ diagram (Figure \ref{figV1309Sco-OMCresid}).

Finally, we consider the dependence of the exponential formula on its variables, $P_0$, $\tau$, $t_0$.
The {\it shape} of the exponential formula in \OMC\ or period plots is fixed except for scaling.
The parameter $P_0$ sets the scale of the vertical axis, while $\tau$ sets the scale of the time axis.
The time of merger, $t_0$, shifts the model horizontally.
This fixed shape stands in sharp contrast to timing models based on third body-induced time delays.
The numerous degrees of freedom in those models can be used to fit a wide variety of shapes.
Hence the distinctive shape of the exponential formula is a strong test of the merging stars hypothesis.

\section{KIC 9832227 PHOTOMETRIC OBSERVATIONS}\label{secKIC-phot}

Me15 reported that the timing data of KIC 9832227 from 1999--2014 were well fit by an exponential formula.
We present here the details of the data used, the processing, and the modelling.
We also present data from 2015--2016, showing that they follow the extrapolation of the previously established exponential formula.  We also fit revised parameters to the exponential formula using all of the data.

\subsection{The data sets}\label{subKIC-photdata}

The Northern Sky Variability Survey (NSVS) monitored KIC 9832227 from 1999 April to 2000 March \citep{2004AJ....127.2436W}.
It appeared under two object identifications: 5597755 and 5620022, with 238 and 217 unfiltered measurements respectively as it straddled the boundary of two program fields.
Median measurement uncertainties were 0.034 and 0.044 mag, respectively.
We converted the times given in Julian Date (JD) to barycentric Julian Date (BJD) using the Python package astropy and the DE423 solar system ephemeris from JPL\footnote{URL: http://docs.astropy.org/en/stable/coordinates/solarsystem.html}.
%\bf next line
(In this paper we follow the convention for modified Julian Date used in the Kepler Eclipsing Binary Catalog \citep{2016AJ....151...68K}: ${\rm MJD} \equiv{\rm JD} - 2400000.0.$)
To insure a common zero point for the two data sets we computed a Fourier series fit to the average light curve folded on the exponential ephemeris of Me15 and computed the average residual for each data set.
We subtracted 0.062 magnitudes from the 5620022 data to place them on the same scale as the 5597755 data.
We clipped four data points that had residuals relative to this average light curve greater than 0.15 mag.
We iterated these calculations so that the fit used in the end reflected both the zero point correction and the editing.

The All Sky Automated Survey (ASAS) monitored KIC 9832227 from 2006 May to 2008 January \citep{2009AcA....59...33P}.
It appeared under the identification 192916+4637.3 for the V filter data and 192916+4637.4 for the I filter data, with 83 and 104 measurements, respectively.
ASAS computed light curves using several aperture widths.
We followed the choices they made for the figures on their web page (number 2 for the V filter and number 0 aperture for the I filter).
We converted the times given in Heliocentric Julian Date (HJD) to JD and then to BJD.  
While the two data sets spanned approximately the same time range, the I filter data were mostly taken near the end of the time range.
As with the NSVS data, we sought highly discrepant measurements and clipped two from the V data set and one from the I data set.

The Wide Angle Search for Planets (WASP) survey monitored KIC 9832227 intensively from 2007 May to 2008 August \citep{2010A&A...520L..10B} and used the identification 1SWASP J192916.01+463719.9.
These data were obtained with a broadband filter with a passband from 400 to 700 nm.
We converted the times given in JD to BJD.
Data were taken in four observing seasons (spanning 2004 May to 2008 Aug), but the first two seasons had too few measurements to establish an orbital phase.
The third season was of generally good quality except for one night (JD 2454234) which had great scatter and which we did not use.
Again, we sought highly discrepant measurements, setting the limit at 0.1 magnitudes, which clipped 58 data, leaving 13,298 good measurements in two observing seasons.

The Kepler spacecraft fortuitously observed KIC 9832227 during its primary mission from 2009 May to 2013 January.
Originally, this target was classified by the Kepler binary star working group, who classified it as an eclipsing binary star \citep{2011AJ....141...83P}.
Separately, this target was granted continued follow up Kepler photometry through Guest Observer Director's Discretionary time, under proposal GO program \#30024 during quarters 10--13.
For all the quarters, this target was regularly observed in long cadence mode (values averaged in 29.4 minute bins with typical brightness uncertainties of 0.07 mmag) for a total of 53,410 measurements.
The Kepler passband spanned 423--897 nm.
Details of the Kepler data processing and calibration for this target are provided in \citet{2016AJ....151...68K} and references therein. 
Kepler information on KIC 9832227 can be found in the Kepler Eclipsing Binary Catalog\footnote{http://archive.stsci.edu/kepler/eclipsing\textunderscore binaries.html} or at the MAST archive\footnote{http://archive.stsci.edu/kepler/data\textunderscore search/search.php}.

KIC 9832227 was observed between 2013 June and 2016 September with Calvin College's two identical (0.4-m OGS Ritchey-Chr\'{e}tien) telescopes: one operated remotely in Rehoboth, NM, at an elevation of 2024 m, and a second located on the Calvin campus in Grand Rapids, MI, at an elevation of 242 m. The Rehoboth telescope has an SBIG ST-10XE camera with a plate scale of 1.97 arcseconds per pixel (binned 2x2), while the Grand Rapids telescope has an SBIG ST-8XE camera with a plate scale of 1.58 arcseconds per pixel.  Given better viewing conditions the New Mexico data have lower noise, but an average anticorrelation in weather conditions between the sites is advantageous for a monitoring program.  Altogether 32,421 images were obtained on 165 distinct nights: 23 in Grand Rapids, 143 in Rehoboth.

Red filters were used every night to monitor timing, approximating the Kepler passband: Bessel R on 148 nights, Sloan $r$ on 17 nights, 28,289 red images in total.
For nights during which only the red filter was used, images were obtained every 61 s in New Mexico (every 130 s in Michigan).
In addition Bessel B was used on 9 nights early on in order to establish color change with phase.
These data ruled out the possibility suggested by \citet{2013arXiv1310.0544K} that KIC 9832227 might be a pulsing star with an unusual light curve.
Bessel V and I images were also obtained on 12 nights to test the need for altitude dependent color corrections.
We used MaxIm to perform differential photometry of the target relative to the average brightness of four reference stars, chosen for proximity in brightness, position, and color, and hence found no need for color correction.
Nightly average brightness over the four years for each reference star brightness were examined to confirm lack of significant variability.
The standard deviation of nightly averages for individual reference stars ranged from 3 to 7 mmag.
The median random uncertainty on individual measurements of KIC 9832227 was 5 mmag for data taken in New Mexico and 7 mmag for data taken in Grand Rapids. 
The reference stars used were KIC 9771791, KIC 9832338, KIC 9832411, and GSC 03543:01558.
The aperture had a radius of 5 pixels, the gap a width 2 pixels, and the sky annulus a width of 3 pixels.
Finally, on 11 of the nights with Sloan $r$, we also used Sloan $g$, $i$, and $z$ filters in order to determine average colors and a photometric temperature (see \S \ref{subsecTemp}).

As a complete light curve cannot be obtained in a single night, we group the observing nights into 38 epochs for purposes of timing analysis.
Table \ref{tabCalvinLog} lists each epoch name, along with the first and last night included, the number of nights observations were made, which observing site(s) was used, how many images were obtained through the R or $r$ filters, what other filters were used, and whether coverage of all orbital phases was complete.
Figure \ref{figCalvinLCs} shows representative light curves from five epochs spanning 2016 May to July.
Successive epoch are offset by 0.1 mag for clarity.
Epochs with complete orbital phase coverage are plotted in red, while incomplete ones are gray.

\subsection{Timing Analysis of the Kepler Data: Discovery of Component C}\label{subsecKIC-OMC}

The Kepler data set is by far the most precise and continuous of the data sets.
Hence it is the natural choice to use for a cross correlation template.
It is also valuable to analyze for short timescale variations.
The results inform the choice of analysis tools and limitations of interpretation.

Figure \ref{figKeplerLC} shows in red a 10th order Fourier fit to all of the Kepler magnitudes along with black dots for the individual measurements (all folded on the exponential ephemeris given in \S \ref{subsecExp}).
The variation of the average light curve spans 0.20 mag.
The wide spread of the data around the average is due to real changes in the light curve shape as described by \citet{2013arXiv1310.0544K}.
Figure \ref{figSampleLCs} illustrates the nature of the variations by showing data from five consecutive orbits beginning on MJD 54983 in red and comparing them to data beginning on MJD 55118 in blue.
The asymmetries between the maxima are reversed, as are the asymmetries between the minima.
We interpret these asymmetries as due to starspot activity.

The strength and arrangements of spots gradually vary with time.
At times a single starspot region dominates and can be seen drifting in longitude.
Similar behavior is seen in the Calvin College data.
This is illustrated in Figure \ref{figCalvinResids} which shows the residual of the data from Figure \ref{figCalvinLCs} relative to the average Kepler light curve (Figure \ref{figKeplerLC}).
These residuals are characterized by a variation of 0.05 mag with a single maximum that drifts in two months from orbital phase 0.3 in the top light curve to 0.75 in the bottom one.
We defer a detailed analysis of starspots to a later publication, in which we will explore variation of the drift rate over the years.
The relevance to this study of orbital timing is that large and time variable starspots can introduce an additional signal into timing determinations, an effect we wish to eliminate.

\citet{2014AJ....147...45C} performed an analysis of eclipse timing of the 1279 close binaries in the Kepler data set.
Close binaries were defined as those with \citet{2012AJ....143..123M} morphology parameters exceeding 0.5.
They computed \OMC\ (for which they use the term ``eclipse timing variations'' or ETV) separately for primary and secondary eclipses using a chain of polynomials to identify ``knots" (essentially points of ingress and egress).
Their (online) plot of KIC 9832227 shows the timing of the two eclipses varying over a range of 35 minutes, largely out of phase with each other, a consequence of drifting and changing starspots.
They also plot the timing of the average of the two eclipses, which reduces the range of variation although there remain abrupt features in which the timing may go up (or down) by five minutes and return over the course of just one or two months.
They identified 236 systems with an oscillatory signal in the timing consistent with light travel delays due to a third body.
Of these, they listed 35 systems identified as having a {\it likely} signature of a third body (which they restrict to third body periods less than 700 days so that two full cycles were covered by the data).
They also listed 31 systems better fit with a constant period derivative (i.e., \OMC\ plots consistent with a constant, nonzero period derivative).
Lacking an obvious overall trend in the plot, they did not include KIC 9832227 in either list.

We note in passing that we checked whether there are contact systems in the Kepler catalog that exceed the threshold $\dot P_t$ that we set using OGLE data.
The Kepler data set has a much shorter time baseline than the OGLE data, so the constant period derivatives found may not represent sustained values as they may turn out to be portions of third body oscillations with longer periods.
We downloaded the Kepler data for the 31 systems \citet{2014AJ....147...45C} listed as having a constant period derivative and computed the values of the derivatives.
Of those systems that they also classified as contact binaries, the only system with a period derivative of either sign with a magnitude exceeding $\dot P_t$ was KIC 9840412 for which $\dot P$ was $-3.3\times 10^{-8}$. 
We observed this system for two nights in 2016 July (using the equipment described in section \ref{subKIC-photdata}) and found the observed orbital phase $19.3\pm0.5$ min behind the phase expected, indicating the magnitude of $\dot P$ for this system has subsequently decreased.
Hence, the Kepler data set contains no exceptions that would require adjustment of $\dot P_t$ to a more extreme value.

For our first pass evaluating the timing of the Kepler data of KIC 9832227, we made a template like that of Figure \ref{figKeplerLC}, but folded on the linear ephemeris of \citet{2014AJ....147...45C}.
We then performed a set of cross correlations to determine the \OMC\ for segments of the data spanning 5 consecutive orbits.
While optimal timing of wide binaries is done by analysis of eclipses, cross correlation makes sense for contact binaries as the light is continuously variable.
Furthermore, similar to the averaging done by Conroy et al. of primary and secondary eclipse times, cross correlation averages over the entire cycle and removes the first order timing distortions of starspots.
This is plotted with blue circles in Figure \ref{figConroy}a, and looks very similar to the average \OMC\ figure of Conroy et al. 
We connect adjoining measurements with lines to emphasize that observational noise is negligible in this plot.
The few month long excursions of up to five minutes are resolved in time, with a series of independent measurements that proceed smoothly.
These excursions are residual effects of changing starspot patterns.
We also plot a quadratic fit to the cross correlations.
This shows a long term trend with a negative period derivative, $-5.4\times 10^{-9}$, among the most extreme contact binary period derivatives found by Conroy et al.

We wish to remove the excursions as much as possible in order to see the underlying trend more clearly.
We explore the potential for doing this with Fourier components by plotting the amplitude of the first six components versus time for 5 orbit data segments (Figure \ref{figFALC}a).
The modulation due to orbital geometry is principally in the even numbered orders, plotted with solid lines.
These are stronger and less variable than the odd numbered orders (plotted with dashed lines), which are largely due to the time variable spots.
We also plot the amplitudes of the Fourier components of the residual of the data to the folded average light curve (Figure \ref{figFALC}b), thereby removing the contribution from orbital geometry.
Since light modulation of starspots is a consequence of rotation (and therefore gradual), we expect most of the power in the first order, with progressively less power for increasing orders, as Figure \ref{figFALC}b shows.

Hence use of even numbered Fourier orders for timing should suppress the starspot contribution.
The best order to use will be a tradeoff between the improved separation of signals at higher orders and the increased observational noise at higher orders.
We plot the second order determination in red with squares on Figure \ref{figConroy} along with the cross correlations.
The similarity of shape indicates this order is no better at separation than the cross correlation.
The vertical offset indicates a 3 minute error in the time of zero phase reported by Conroy et al.

In Figure \ref{figConroy}b we plot the second, fourth, and sixth orders with five minute offsets inserted between them so that the orders may be seen separately.
The short-timescale excursions seen in the second order are only half as big in the fourth order and are essentially gone in the sixth order.
The observational scatter which is invisible in the second order becomes significant by the sixth order.
Nonetheless the sixth order is the cleanest indication of long term trends.
In addition to the general curvature seen in the cross correlation, there is also a broad dip around 2011.

For our second pass, we plot the sixth order again (Figure \ref{figSmallBodyOMC}), but this time relative to the exponential ephemeris of \S \ref{subsecExp}.
For the Kepler epoch, this essentially means removing a period derivative of $-3.4\times 10^{-9}$.
The parameters of that ephemeris are determined by comparison of the Kepler timing to earlier and later measurements.
Hence the relatively flat result is an indication that the period and period derivative of the exponential ephemeris apply within the Kepler data.

This Figure also shows a small oscillation of the \OMC\ around the average, which we fit with a simple sine wave of semiamplitude $1.69\pm0.03$ minutes and a period of $590\pm8$ days and plot as a blue line in Figure \ref{figSmallBodyOMC}.
\citet{2014AJ....147...45C} interpreted \OMC\ diagrams like this as caused by light travel time variation arising from the orbit of the binary around a third body.
Although the time signature of a third body may be more complex than a sine wave (for nonzero eccentricity) the data are not of adequate quality to justify a more complex fit.
We designate this third body component C.
Assuming a binary mass of 1.71 $M_\odot$ (as determined in \S \ref{subsecBM}), then $m_C\sin i_C$ is 0.11 $M_\odot$, where $m_C$ is the mass of component C and $i_C$ is the inclination of its orbit to our line of sight.
Finally, our best estimate of the time of zero orbital phase in the Kepler data set comes from setting the average of the sine wave to zero: BJD$_0$ = MJD 55688.49913(2).

\subsection{Exponential Model of Orbital Timing}\label{subsecExp}

%\bf next lines
Me15 fit the KIC 9832227 timing data taken through 2014 to the exponential formula of Equation \ref{eqnExp} and found the best fit parameters to be $(P_0,\tau,t_0)$ = (0.4579603 d, 0.089 d, 2020.8(15)).
Updating this fit using data through 2016 September we find $(P_0,\tau,t_0)$ = (0.4579615(9) d, 0.114(14) d, 2022.2(7)).
The numbers in parentheses indicate the uncertainty on the final digit(s).
The new values are consistent with the previous ones to within their uncertainties.
The most relevant change is the reduction in the uncertainty on $t_0$, which is sensitive to data taken near $t_0$.
In this section we present the details of the fitting and plot the results in several ways to give deeper insight.

Figure \ref{figKIC-OMCresid} shows the residuals of all of the orbital timing data relative to our new exponential ephemeris and the BJD$_0$ from the preceding section.
The three pre-Kepler data sets are relatively sparse, so a single cross correlation was done for each observing season.
The horizontal error bars indicate the time range covered for each timing measurement.
An advantage of averaging over data randomly spread over a long time range is that it should also average out residual effects of variable starspots.
The Kepler timing estimates come from the sixth order phases shown in Figure \ref{figSmallBodyOMC}, but averaged in groups of seven to reduce the crowding.
The Calvin estimates use second order phases rather than 6th order as there is less random noise.
As a consequence, these values are precisely measured, but may contain starspot-induced excursions like those seen in the 2nd order Kepler data (Figure \ref{figConroy}b).
The 25 epochs with complete phase coverage were used in the fit.
The 13 epochs without complete phase coverage show similar values but are omitted as systematic effects of  starspots are greater in them.
The effect of component C exceeds the measurement uncertainties in both the Kepler and Calvin data, and the residual effect of starspots exceeds the measurement uncertainties in the Calvin data.
These two effects should have no systematic influence on the exponential fit parameters as they are small in amplitude compared to the long term drift in the timing and oscillate on short time scales.
However, they do increase the effective measurement noise.
Hence the uncertainties used in the fit for the Kepler and Calvin data sets were based on their respective residual scatter.

Figure \ref{figKIC-OMC} shows the timing data relative to a constant reference period, the standard \OMC\ plot.
As with V1309 Sco (Figure \ref{figV1309Sco-OMC}) the reference period was chosen to be the actual period near the middle of the data set.
The sharper curvature in later years than in earlier years is again the mark of an increasingly negative period derivative with time.
The exponential fit of Me15 is marked by a black dashed line, while our revised fit is given as a solid purple line.
The 18 new Calvin data points from 2015--2016 (filled purple circles) follow the prediction of the exponential line.
As forecast by Me15 the magnitude of the period derivative exceeded $\dot P_t$ in 2015.
Unlike the case of KIC 9840412 (described in the preceding section), there is no evidence to date of a reversal in the timing trend.
After presentation of our spectroscopic data in \S \ref{secSpec}, we will discuss in detail the limits that can be placed on alternative interpretations of the long term trend in timing as light travel time delay (\S \ref{subsecCompD}).

Although the exponential model parameters were fit directly to the \OMC\ timing data, it is instructive to plot period versus time as well to see directly the amount of change in the period.
Figure \ref{figKIC-P} shows the exponential formula period with time along with three period determinations made by separately fitting the slope of the \OMC\ plot for pre-Kepler, Kepler, and Calvin data, respectively.
As before, uncertainties in the Kepler and Calvin data are determined using the actual scatter in the timing data.
The points are plotted at the times of the midpoint of the time ranges of the respective data sets.
The periods thus derived, listed in Table \ref{tabKIC-P} and plotted in Figure \ref{figKIC-P}, show very significant changes relative to the uncertainties.
It is likewise instructive to plot the period derivative with time, Figure \ref{figKIC-Pdot}, to see directly the change in this quantity.
Two estimates of the period derivative are plotted that are based on the differences in the periods in the preceding plot.
These are marked at the times midway between the times of the period estimates.
The apparent disagreement with the model curve is an artifact of the curvature of the model over the time range used to make the estimate.
The increasingly negative period derivative with time is the essential signature of the merging star model.

\section{KIC 9832227 SPECTROSCOPY}\label{secSpec}
Spectroscopic data were obtained for KIC 9832227 to supplement the photometric datasets.
The spectra provided the stellar characteristics of the system (e.g. effective temperature, log g, and metallicity) and the velocities of the stellar components.
High resolution spectra were obtained from two sites: Apache Point Observatory (APO) and the Wyoming Infrared Observatory (WIRO).

At APO, spectra were obtained from 2012 July until 2015 June using the ARCES echelle spectrograph on the 3.5~m telescope.
The wavelength coverage is from 320~nm to 1000~nm, and the spectrograph has a resolution of R $\sim$ 31,500.
Our exposure times ranged from 900 to 1800 s, achieving a S/N of 20 and 30, respectively.
Table \ref{tabspecobs} lists the timing details of the observations.

For the spectra, we used a subset of wavelength coverage from 460--680~nm due to prohibitively low S/N at both the red and blue ends of the spectra.
We also avoided regions where atmospheric telluric lines contaminated the spectra.
This wavelength region was also used for analysis of the WIRO spectra.
The APO spectra were reduced using the standard IRAF data reduction techniques.
For calibration purposes, additional radial velocity and spectrophotometric standard stars were observed on the same nights as KIC 9832227.
We used radial velocity standards HIP 095446 and HD 196850.
For spectrophotometry, the standards Feige 34 and 66 were observed.

Using the long-slit spectrograph on the WIRO 2.3~m telescope, data were obtained over eight nights in 2015 June with the 1800 mm$^{-1}$ grating, which provided wavelength coverage over 540--670~nm at a dispersion of 0.63 {\AA} pixel$^{-1}$.
Typical integration times of 600~s resulted in spectra with S/N $\thicksim$100.
Reductions followed the standard procedures, and CuAr arc lamp exposures were used for wavelength calibration to $\thicksim 0.0015$~nm rms.
The calibrated spectra were then normalized using line-free regions and a Heliocentric velocity correction was applied.
High S/N spectra for three radial velocity standard stars (HIP 095446, HIP 96514, and HD 196850) were also obtained.

\section{ANALYSIS}\label{secAnalysis}
\subsection{Surface Temperature and Metallicity}\label{subsecTemp}
\subsubsection{Photometric Estimates}\label{subsubsecTempPhot}
The Kepler Input Catalog (KIC) gives a temperature of $5854\pm100$~K, based principally on a set of Sloan colors $(g-r, r-i,i-z)=(0.454,0.099,0.029)$ mag \citep{2012ApJS..199...30P}.
\citet{2014MNRAS.437.3473A} used these colors along with ultraviolet and infrared measurements to estimate the primary and secondary temperatures separately: $(T_1,T_2)=(5912\pm360,5930\pm560)$~K.
Given the time variability of the system due to orbital variation as well as varying starspot strength, it is worthwhile to establish time-averaged colors and the degree of color variation over an orbit.

During 2014 October 25--31 UT we observed the field alternately in $g$, $r$, $i$, and $z$, completing 196 sets of measurements covering all orbital phases.
Similarly in 2014 November 4--9 UT we made 141 sets of measurements covering all orbital phases.
We calibrated our magnitudes using KIC colors for five reference stars chosen for proximity on the sky, a narrow color rang that brackets the color of KIC 9832227, and lack of significant time variability: KIC 9771402, 9771453, 9771791, 9832338, and 9832411.
Nightly averages of the three Sloan colors of the reference stars typically varied by 1 mmag from night to night, with a maximum variation of 3 mmag.
For KIC 9832227, we fit a sixth order Fourier model to the light curve for each filter for each epoch.
The best estimate of orbital average magnitude is the constant term of these fits.
The difference in these averages from October to November was $<$1.5~mmag, consistent with random error.
The filter averages of the two months were $(g,r,i,z)=(12.721,12.279,12.171,12.157)$~mag.
The color averages of the two months were $(g-r,r-i,i-z)=(0.442,0.108,0.014)$~mag with random errors less than 1 mmag.
Variations from these colors with orbital phase were less than 9 mmag in either month.

The differences between the KIC colors and our average values, $(\Delta (g-r),\Delta (r-i),\Delta (i-z))=(0.013,-0.009,0.018)$~mag, are greater than the orbital variation we observed.
Nonetheless, the differences are not enough to affect the temperature estimate beyond the 100~K uncertainty.

\subsubsection{Spectroscopic Determination}\label{subsubsecTempSpec}
Spectra were analyzed for $T_{\rm eff}$ and [Fe/H] determination using the \emph{iSpec} Python tool developed 
by \citet{2014ASInC..11...85B}. This tool provides a number of different line databases and atmospheric models to 
allow the user to determine atmospheric parameters of the input stellar spectrum by fixing input 
parameters that are well known while fitting for unknown parameters. The tool also allows for determination 
of radial velocity by cross-correlating the data against any of a number of template spectra. Procedures were
generally followed as described by \cite{2014A&A...569A.111B}, the \emph{iSpec} user manual, and S. 
Blanco-Cuaresma (2015, private communication).

The normalization of the APO spectra was checked over the reduced wavelength range and adjusted as needed. Normalizing 
the extracted orders of echelle spectra is notoriously challenging, particularly in low-S/N spectra, and in some 
instances small adjustments to the IRAF normalization were needed.
The renormalization was done with \emph{iSpec} using several spline and low-order polynomial fitting models.
These were repeated to identify systematic effects on the final parameter fit values.

APO spectra from orbital phase $\phi$ = 0.0 were co-added into a master $\phi$ = 0.0 APO spectrum with maximum S/N. The 
same was done for all phase $\phi$ = 0.5 APO spectra, as well as WIRO spectra from these phases. 
These four master spectra were fit using the MARCS model atmosphere of
\citet{2008A&A...486..951G}, solar abundances from \citet{2007SSRv..130..105G}, and the most recent atomic 
line list including hyper-fine structure transitions \citep{2015PhyS...90e4010H} from the \emph{Gaia} ESO 
survey \citep{2012Msngr.147...25G}.
Due to the high rotational speed of this contact system, spectral features were 
Doppler-broadened ($v\sin i\thicksim 150$~\kms ) to the extent where the data were not sensitive to macroturbulence and microturbulence, so default values of 0.0~\kms \ and 1.0~\kms , respectively, were adopted.
Small increases in these parameters did not affect the results.
A precise surface gravity also could not be determined, and so $\log~g$ was fixed at 4.0.
Results were consistent with tests using 4.5 as well.

Within \emph{iSpec} the initial value of $T_{\rm eff}$ was varied systematically between 5800~K and 5900~K (spanning the KIC value) to check for consistency in the output fit values, while initial values for [Fe/H] were varied between $-$0.05 and +0.05.
The best fit spectra for phases 0.0 and 0.5 are shown 
in Figures \ref{figspecapo00} and \ref{figspecapo05}, where the surface temperature and 
metallicity for the system were determined to be $T_{\rm eff}$ = 5828 $\pm$ 42~K (random) $\pm$ 40~K (systematic) 
and [Fe/H] = $-$0.04 $\pm$ 0.03 (random) $\pm$ 0.04 (systematic). The largest source of systematic uncertainty was the
renormalization of the echelle spectra, while the random error is a combination of fitting error reported by \emph{iSpec}
and the scatter in the output fit values reported while the parameters were being incrementally varied.

Throughout this work we use linear limb darkening coefficients from \citet{1993AJ....106.2096V}\footnote{as updated at http://faculty.fiu.edu/$\thicksim$vanhamme/limb-darkening}, using $\log g=4.2$ (as found for the primary star in \S \ref{subsecBM}), the temperature and metallicity given above, and the relevant filter or wavelength range.
For the APO and WIRO spectra, the coefficients were 0.55 and 0.511, respectively.

\subsection{Binary Star Velocities}\label{subsecVel}

An effective way to digest the velocity information in a spectrum is to compute the broadening function, a least-squares deconvolution of the data that represents spectral power as a function of radial velocity.
\citet{2004IAUS..215...17R} reviewed the advantages of this approach over the simpler cross-correlation function.
Like the cross correlation, this technique requires a template spectrum for comparison.
Fortunately, in the case of a contact binary system both components have very similar atmospheric temperature and surface gravity, so one template works equally well for both stars.
\citet{2005A&A...443..735C} published a grid of model atmospheres, from which we selected the one with solar abundances and ($T,\log g$)=(5750~K,4.0).
To compute broadening functions, we selected spectral regions with relatively good signal-to-noise ratios and free from strong terrestrial lines.
For the APO data we used the wavelength range 373.0--686.6~nm, smoothed with a 3.5~\kms \ Gaussian with results in 3.5~\kms \ bins.
For the WIRO data we used the wavelength range 539.5--669.4~nm, with results in 7.0~\kms \ bins.
We made a sequence of three analysis passes.

\subsubsection{First Pass: Fitting Gaussians to Individual Spectra}\label{subsubsecPass1}
The first pass follows the typical application of the broadening function \citep[e.g.,][]{2002AJ....124.1746R}.
A pair of Gaussians are fit individually to each spectrum.
Then the Gaussian center velocities as a function of orbital phase are fit to find the radial velocity semiamplitudes of the two stars, $K_A$ and $K_B$, and the Heliocentric radial velocity of the system barycenter, $\gamma$, where the subscripts A and B refer to the primary and secondary stars, respectively (components A and B).
Figure \ref{figbfrvcurvegaussian} shows the radial velocity curves for each component.
The sinusoidal fits shown yield ($K_{A}$,$K_B$,$\gamma$) = (48.9,226.6,$-22.2\pm 0.8$)~\kms .
This implies a stellar mass ratio $m_B/m_A=K_A/K_B=0.216$.

\subsubsection{Second Pass: Fitting Gaussians to the Combined Data Set}\label{subsubsecPass2}
A statistically better approach to the fitting is to do a global solution for free parameters.
For Gaussian fits, there are the three parameters of interest ($K_{A}$,$K_B$,$\gamma$) along with the strengths and widths of the two peaks, a total of seven parameters.
This is far less than the number of free parameters in the first pass: six model parameters (strength, width, and velocity of each star) times the number of spectra solved.
To be conservative, we also omitted from the fit spectra taken within 0.1 orbital phase of an eclipse to avoid systematic effects from partial eclipses.
The eclipses in this system are shallow, however, and comparison showed this restriction had negligible effect on the fit (and indeed that the broadening function model fit these data well anyway).
Since the data are taken under varying conditions, it is also worthwhile to weight their contribution to the final fit.
We determined the weights from the residuals to the fits, iterating until the fits and weights converged.
As a consistency check, we performed the fit separately for the APO and the WIRO data.
The APO fit yields ($K_{A}$,$K_B$,$\gamma$) = (46.0,224.0,$-$24.3)~\kms ,
while the WIRO fit was (47.4,221.9,$-$22.3)~\kms .

\subsubsection{Final Pass: Fitting Rotationally Broadened Spheres to the Combined Data Set}\label{subsubsecPass3}
A better approach still is to build on the second pass approach by also using a physically motivated model: two rotating spheres.
For a synchronously rotating contact binary, rotational broadening is comparable to the orbital velocity and larger than other sources of line broadening.
Since the wing of a rotationally broadened line cuts off more precipitously than that of a Gaussian broadened line, forcing a Gaussian shape will systematically shift the measured radial velocity of the smaller star.
The fitting parameters we use now are $K_{A}$, $K_B$, $\gamma$, $v_A\sin i$, $v_B\sin i$, $R$, and $A$,
where the new parameters are the projected stellar rotational velocities, the ratio of the integrated signal from the second star to that of the first, and the sum of the integrated signal from both stars.
In addition to this approach providing a better fit, three of the new parameters (the rotational velocities and the signal ratio) are physically meaningful values that will be used as a consistency check when modelling the system geometry.
A final element of the model is convolution with a fixed Gaussian (to account for instrumental effects).
We determined the appropriate Gaussian widths by fitting Gaussians to our standard star spectra: 5.5~\kms\ and 31.7~\kms\ for APO and WIRO, respectively.
These values are consistent with the combined effects of instrumental resolution and smoothing.
The model spheres included limb darkening, with coefficients 0.629 and 0.529 for APO and WIRO, respectively.

To exemplify the results Figure \ref{figBFFit} shows model fits to one APO and one WIRO broadening function.
The orbital phases are labelled in the upper right corner.

The final APO fit yields ($K_{A}$,$K_B$,$\gamma$,$v_A\sin i$,$v_B\sin i$,$R$) = (49.1,215.5,$-$22.9,152.0,88.9,0.304),
while the WIRO fit was (49.8,216.0,$-$22.2,150.3,96.7,0.311). (All units are \kms\ except for the dimensionless ratio.)
The residuals to this model were 12\%\ smaller than the second pass residuals for the APO data (10\%\ for the WIRO data), indicating a significantly better model shape.
Likewise, the agreement between the two data sets was better than for the Gaussian model, consistent with elimination of a systematic problem.
The lone exception is the rotational velocity of the secondary star.
For this we conclude the WIRO data (which has lower spectral resolution) is not sensitive to this parameter.
Otherwise, the differences between the two fits are consistent with random uncertainties on fit parameters.
In \S \ref{subsecBM} we will see the rotational velocities and the ratio parameter are also very consistent with the geometric model of all the data.

The most important result of the velocity study is the mass ratio, which is 0.228/0.231 for the APO/WIRO data.
These values are $\thicksim$6\%\ greater than our first pass value.

\subsection{Physical Parameters of the Binary System}\label{subsecBM}

In Table \ref{tabPhysicalParameters} we summarize the physical parameters of KIC 9832227 as inferred from the time average Kepler light curve (Figure \ref{figKeplerLC}) and the spectroscopic temperature and radial velocities.
First, we used the APO mass ratio in a fit of the light curve made with Binary Maker 3.0 \citep{2002AAS...201.7502B} to find the system inclination ($i$), fillout parameter ($f$), temperature ratio ($T_B/T_A$), and the parameters of one starspot.
Fixed input parameters include 0.549 for the limb darkening coefficient, 0.32 for the gravity brightening coefficient (appropriate for convective atmospheres \citet{1967ZA.....65...89L}), and 0.5 for the reflection coefficient (again appropriate for convective atmospheres \citet{1969AcA....19..245R}).
As will be shown in \S \ref{subsecMC}, component C is so small that no spectral lines can be detected from it.
Accordingly, we do not expect its continuum emission to significantly affect this light curve analysis. 

The inclination is largely determined by the depth of the orbital modulation.
The fillout parameter is a convenient equivalent of the surface equipotential scaled such that it is zero for a system precisely filling its Roche lobe and unity for a system reaching the outer critical Roche potential.
It is determined by how rounded the eclipses are.
The value, 0.43, indicates an intermediate depth of contact between the two stars.
The temperature ratio is determined by the relative depths of the eclipses.
Temperature ratios exceeding unity (1.021 in this case) are typical of contact binaries with shorter orbital periods.
Note that we have chosen orbital phase zero to be the point at which the secondary star eclipses the primary one, although due to the temperature ratio this is {\it not} the deeper of the two eclipses.
The need for a starspot is evident from the asymmetric maxima.
It is not generally possible to determine a unique starspot configuration from the light curve.
We fit a single spot on the primary star with radius of 9\fdg 6 and colatitude 60\degr .
We solved for the spot longitude (276\degr ) and fractional temperature ($T_{\rm spot}/T_A$=0.825).
These are physically reasonable values and they yielded an excellent fit to the light curve shape.

%\bf next paragraph
The difference between the time-averaged data and the best fit model implies a standard error on the data of only 0.0015 mag.
The uncertainties given for the Binary Maker parameters ($i$, $f$, and $T_B/T_A$) are based on the assumption of random errors of this size.
As no correction was made for the small smoothing effect of the 29.4 minute binning of the Kepler data, there are also systematic effects that are probably comparable to the random uncertainties.

Once the inclination is known, the radial velocities and the orbital period can be used to find the total mass, the semimajor axis ($a$), and the individual stellar masses.
The semimajor axis in turn sets the scale for stellar size.
Since the stars are not spherical, several distinct equivalent radii may be of use.
The radius ($R_V$) in the table is the radius of a sphere with the same volume as the star.
The surface gravity ($g$) makes use of this radius.
The rotational velocity ($v\sin i$) in the table makes use of the back radius (the distance from the star's center of mass to the equatorial point furthest from its companion.
The APO model fit rotational velocities lie between these values and the values one would obtain measuring radius from the center of mass to the inner Lagrange point, a confirmation of the self-consistency of the physical parameters.
The area ratio is the ratio of the surface areas of the two stars.
The individual stellar temperatures are determined in an iterative process such that the brightness weighted average of the temperatures is the average temperature determined from spectroscopy.
The light ratio is the ratio of stellar brightnesses taking into account both the different surface areas and temperatures.
This is within 1\%\ of the independently determined ratio from the APO spectroscopy, another confirmation of the self-consistency of the physical parameters.

The distance was estimated starting from the absolute magnitude of the Sun in each of the four Sloan filters \citep{2010MNRAS.404.1215H}, scaling for the Planck intensities at those wavelengths and for the surface areas, and comparing them to our measured maximum brightness in those filters.
This approach requires a correction for interstellar extinction.
Maps of Galactic extinction provide an upper limit to the extinction as our target is within the Galaxy.
We estimated interstellar extinction by varying E(B--V) until the dereddened colors corresponded with our spectroscopic temperature using the color-temperature relations of \citet{2013ApJ...771...40B} and the extinction curve of \citet{2013MNRAS.430.2188Y}.

%\bf next line
In the table the uncertainties in the temperatures include the random and systematic uncertainties added in quadrature. In general uncertainties on all table quantities were computed by propagating the uncertainties of the two radial velocities and the three Binary Maker parameters as appropriate for each quantity.

%\bf next line
For completeness the table also includes the time of zero orbital phase from \S \ref{subsecKIC-OMC} and the parameters of the exponential timing fit from \S \ref{subsecExp}. 
We conclude the table with ICRS and equivalent galactic coordinates from the UCAC4 astrometric catalog, in which the star is designated 684-067807 \citep{2013AJ....145...44Z}.

\subsection{Spectroscopic Upper Limit on Mass of Component C}\label{subsecMC}

In \S \ref{subsecKIC-OMC} we found evidence in the timing data for a component C with orbital period $P_C=590$~d and mass $m_C\sin i_C$ is 0.11 $M_\odot$.
If the orbital plane of component C is more nearly face on the mass would be much greater than 0.11 $M_\odot$.  
We can place an upper limit on the $m_C$ by looking for its signal in the broadening function.

As noted by \citet{2002AJ....124.1746R} it is often easier to see the signature of a third body in a broadening function than that of a close binary as the tidally enforced corotation of the binary components makes their signal broad and short.
For our considerations, we use the empirical main sequence relations of \citet{1991Ap&SS.181..313D} to estimate radius, temperature and luminosity from hypothesized mass.
A star hotter than the binary can be immediately ruled out as they would dominate the photometric colors.
Likewise, a star similar in temperature to the binary would show a strong narrow spike in the broadening functions in Figure \ref{figBFFit}.
On the other hand, there is no hope of seeing the spectral signature of
an $0.11~M_\odot$ star, which would have a luminosity just 0.04\%\ of the binary luminosity.
We will show that stars of 0.5 or 0.6~$M_\odot$, with 1.9 or 3.8\%\ of the binary luminosity, respectively, define the limits of what could be detected in the broadening function.

We take three steps to optimize the narrow, nonvariable signal expected: 1) use the higher resolution APO spectra; 2) average together all of the spectra from the night with the greatest number of images (2015 June 21); and 3) deconvolve with a template the temperature of the hypothesized star (4000 K and 4250 K for 0.5 or 0.6~$M_\odot$, respectively).
The signal from the binary will be smeared in velocity from the time averaging and diminished in amplitude because their spectra differ from the cooler template.
Figure \ref{figThirdBodyBF} shows the broadening functions optimized for 0.5~$M_\odot$ (a) and 0.6~$M_\odot$ (b) as black lines.
The binary stars produce the very broad peaks at the center of each plot as well as the spikes at the right and left edges, artifacts of the temperature mismatch.

If component C were 0.5 or 0.6~$M_\odot$, we would expect to see a spike near the system velocity of the binary in Figure \ref{figThirdBodyBF}.
The question is how the size of the expected spike compares to noise fluctuations in that portion of the broadening function.
We address this question with a simulation.
We add the template to the June 21 data, binning to match the wavelength resolution and scaling the amplitude for the correct relative strength (for the given temperature and stellar surface area) at each wavelength.
No additional noise is added as the Poisson noise of the small additional component is much less than the random noise already present in the data.
The broadening functions of these simulated companions are plotted in Figure \ref{figThirdBodyBF} with gray lines.
The black and gray lines differ only near the velocity of the hypothesized companions.
In both cases, the gray spike far exceeds the random variation of the observed function.
Hence, we conclude $m_C$ is less than 0.5~$M_\odot$, and $i_C$ is greater than 13\degr .

\subsection{Alternative Interpretation of Orbital Timing: Light Travel Delay from a Component D}\label{subsecCompD}

While the long term trend in the timing data of KIC 9832227 remain consistent with an exponential formula,
we cannot consider it a likely interpretation until we can reject the more mundane alternative of light travel delay due to a massive companion star orbiting the KIC 9832227 triple system with a period of 25+ years.
(We designate this hypothetical companion component D.)
This is especially true since the $\dot P$ threshold has only just been reached.
There is nothing to exclude such a body {\it a  priori}, although it is not required here to explain the formation of the close binary as component C is adequate for that.

Unlike the exponential formula, which can only have one shape in the timing plot with two scale factors, the various orbital parameters allow companion orbit models to exhibit a wide range of shapes.
For our calculations, we assume the companion orbital plane is inclined by 90\degr\ (i.e., includes our line of sight, hence maximizing its effect), the mass of the triple is 1.82 $M_\odot$ (using the minimum mass for component C), and the reference period used in the timing plot is the intrinsic binary period.
The latter assumption is for specificity and has negligible affect on our conclusions.
For each assumed companion mass ($m_D$), we fit the orbital period ($P$), eccentricity ($e$) angle between the line of sight and the line of apsides ($\omega$), and time of periastron passage ($t_p$).
These models account for the recent epoch of large period derivative by having an eccentric orbit with the time of periastron near the current date.
They predict an abrupt turnaround in the timing plot in the near future and ultimately periodic behavior in the long run on the orbital period.
Best fit models for $m_D$=0.8, and 1.0 $M_\odot$ are shown as red and green lines in Figure \ref{figKIC-OMC}.
The model parameters,($m_D$,$P$,$e$,$t_p$,$\omega$), are (0.8~$M_\odot$,25.1~y,0.609,43\degr, 2016.5) and (1.0~$M_\odot$,30.7~y,0.569,53\degr, 2017.2).
The 0.8~$M_\odot$ model is a marginal fit, differing by 5 minutes for the most recent data (which could be accounted for by fortuitous starspots).
The 1.0~$M_\odot$ fits the existing data well.

We conclude the minimum mass for a hypothesized component D is 0.8~$M_\odot$.
As discussed in the preceding section, any nondegenerate star of this mass would easily be detected in the broadband colors or the spectroscopic broadening function.
We are unaware of any known degenerate companions to triple systems, so this hypothesis itself is unlikely.

Nonetheless, this alternative model can be tested with additional data.
The timing prediction rapidly diverges from the exponential formula in the next year.
Additionally, the barycenter velocity of the triple system is predicted to rise sharply in the next few years and then level off as the companion moves away from periastron.
Figure \ref{figKIC-RVD} shows the radial velocity of the triple system as it orbits a $1.0~M_\odot$ component D (solid black line).
The radial velocity of component D is shown as a dashed red line. 

\section{DISCUSSION}\label{secDiscussion}
We briefly discuss three aspects of the merger process for which the case of KIC 9832227 may be instructive.

\subsection{Outburst Trigger}\label{subsecTrigger}
While a number of papers have modelled the physics of the observed outburst of V1309 Sco, none cover the full decade preceding the outburst.
Hence there is no physical basis for the exponential formula (Equation \ref{eqnExp}) used to describe the early period changes, nor any insight into what sets the timescale, $\tau$.
Te11 suggest the relevance of both the onset of the Darwin instability and dynamically unstable mass loss.
Whether the tidal viscosity timescale is adequate for the Darwin instability in the case of a contact system has been discussed back and forth by \citet{2014ApJ...788...22P}, \citet{2014ApJ...786...39N}, and \citet{2016RMxAA..52..113K}.
We note here only that the observed mass ratio for KIC 9832227, 0.23, is far from the expected value for the onset of the Darwin instability, 0.1.
As models are developed covering the decade approaching outburst, a wider range of mass ratios should be considered.

\subsection{Outburst Timing}\label{subsect0}

As noted in the introduction, the beginning of the V1309 Sco outburst preceded $t_0$ by 1.8 years.
What might this time difference be in other systems?
If the trigger is a certain rate of energy deposition, the rate of loss of orbital energy scales with $\dot P$.
For fixed $\dot P$, the time offset scales with $(P_0\tau)^{0.5}$ (Equation \ref{eqnTime}).
Alternatively, if the trigger is a certain fraction of orbital energy deposited, it may occur after the same change in $P/P_0$, so the time offset would scale with $\tau$ (Equation \ref{eqnExp}).
At the time of the observed outburst of V1309~Sco, $\dot P \approx -5\times 10^{-5}$ and $P$ was just {2.2\%} less than $P_0$.
Given the much smaller $\tau$ found for KIC~9832227, these two milestones are not reached until a month or so before $t_0$.
Hence for KIC~9832227, we expect $t_0$ to be a good estimate of the actual time of outburst.

We note for context that even the currently observed value of $\dot P\sim -1\times 10^{-8}$ implies a large energy dissipation rate.
As an illustrative calculation, conservative mass transfer from the primary to the secondary at the rate of $1.1\times 10^{-6}~M_\odot~{\rm y}^{-1}$ would produce the observed $\dot P$ and infer a loss of orbital energy at the rate of 19~$L_\odot$, more than five times the observed system luminosity.
As exemplified by the case of V1309 Sco, only a small fraction of the lost energy is emitted from the surface as light.
The primary star might also store the lost energy as increased thermal or gravitational potential energy 

\subsection{Light Curve Transition on Approach to Outburst}\label{subsecTransition}
In the final observing season before merger the shape of the light curve of V1309~Sco changed significantly, with one of the two maxima gradually disappearing and the average system brightness decreasing.
Te11 suggested that as the system approached merger, the primary rotated subsynchronously.
Energy dissipation due to friction occurring near the L1 point would then heat the portion of the primary that would then rotate into view near orbital phase 0.25, increasing the strength of that maximum.
We suggest alternatively that mass lost through the L2 point would accumulate in a disk that would be initially denser on the side viewed at orbital phase 0.25.
This material could cool and obscure the system at this phase, decreasing the strength of that maximum.
Without preoutburst spectroscopy of V1309 Sco, we do not know for sure {\it which} maximum was at orbital phase 0.25, and so the observations do not decide between these two suggestions.
Should a similar transition occur in KIC 9832227, however, we will know how to interpret it.
%\bf next two lines
\citet{1979ApJ...229..223S} show that mass lost through the L2 point escapes from the system for binaries with mass ratios in the range 0.064--0.78 (a range including the value of 0.23 we find for KIC 9832227), resulting in disk-like outflows.
\citet{2016MNRAS.461.2527P} model radiation from such outflows, showing results similar to red nova outbursts can be obtained.

\section{CONCLUSIONS}\label{secConclusions}
\subsection{Summary of Results}\label{subsecSummary}
The central premise of this paper is to use the exponentially decreasing orbital period found by Te11 to precede the red nova outburst of V1309 Sco as a signature to identify the precursor to the next outburst.
The distinctive shape of the exponential formula (fixed except for scale factors) reduces the chance of false positives.
We identify a threshold period derivative ($\dot P_t\approx -1\times 10^{-8}$) that should be exceeded to qualify as a serious precursor candidate.
We estimate our galaxy should have roughly 1--10 observable precursors of events at least as big as the V1309 Sco event (which brightened by 10 magnitudes).

We develop the observational case for the precursor candidate suggested by Me15, KIC 9832227.
The development exemplifies analytical tools needed to identify a precursor in a timely fashion.
Determination of orbital phase versus time (the $O-C$ plot) makes the best use of timing information.
Computations relative to a tentative exponential ephemeris avoid smearing a signal in the presence of a rapidly changing period.
Fourier filtering can help separate the competing signal of time variable starspots (common in contact binaries with shorter periods) from that of true orbital changes.
Spectroscopic analysis of radial velocities is more precise when a global solution is determined and avoids significant systematic errors when a model with realistic rotational broadening is used.

Me15 found that timing data for KIC 9832227 spanning 1999--2014 were consistent with an exponential fit with parameters.
Photometric and spectroscopic observations presented in this paper make two strong tests of the precursor hypothesis in addition to establishing numerous physical parameters of the system.
The most important test passed is that timing data through 2016 September is consistent with the extrapolation of the earlier fit.
Along with this we find $\dot P_t$ has been exceeded as of late 2015.
As a byproduct of the timing analysis we discovered evidence for a small third star (component C with mass 0.11--0.5~$M_\odot$ and $P_C=590$~d).
A companion of this sort is expected as its interaction with the binary may have drawn the binary together in the first place.
The physical parameters of the binary system are listed in Table \ref{tabPhysicalParameters} 
as deduced from a fit to the average light curve and the orbital velocities.
Of particular note is the mass ratio, 0.23, which is inconsistent with the theoretical expectation of 0.1 based on the assumption that the onset of the Darwin instability is the trigger for merger.

The second test passed is that we rule out an alternative interpretation of the long term trend in the timing data in which a luminous companion (a component D) orbits the triple system inducing light travel time delays.
Specifically, we find a component D with a minimum mass of 0.8~$M_\odot$ is required to match the timing data, but that the absence of an observed spectral signature rules out main sequence star 0.5~$M_\odot$ or greater.
The possibility of a degenerate component D remains, although such a system would in itself be a unique discovery.

Passing these two tests greatly strengthens the case for this candidate.
However fortuitous the identification of this precursor candidate seems, the merger hypothesis has had predictive power and we currently have no alternative explanation for its timing behavior.
The prediction of a naked eye nova (2nd magnitude if it brightens as much a V1309 Sco did) in the year $2022.2\pm 0.6$ makes it essential to use the available time to fullest advantage.
A full characterization of the system properties (and perhaps their evolution over the next few years) will provide much more specific constraints of physical models of the outburst process.

\subsection{Further Work}\label{subsecFurtherWork}
Additional work is needed developing the physical theory of mergers, observing the properties of the KIC 9832227 system, and surveying other contact binary systems.

While much progress has been made on the physics of a red nova outburst, no detailed consideration has yet been given to the decade preceding outburst.
This is difficult because of the long time scale, but might be tractable if treated as a series of quasi-steady-state stages.
It is important to develop a physical basis for the exponential formula, and in particular to estimate the timescale as a function of orbital period and mass ratio.
In the present analysis the timescale was fit to the data as the only significant free parameter.
A theoretical constraint on that parameter could be a third significant test of precursor status.
It would also be instructive to develop the theory of the geometric distribution of energy dissipation as merger approaches.
Such a theory could be tested by observation of long term trends in light curve shape and system luminosity.

A variety of additional observations of KIC 9832227 are needed.
First priority is continued monitoring of the orbital timing. 
With each year closer to $t_0$ the prediction of the merger hypothesis becomes more distinctive and hence a more stringent test of the hypothesis.
Along with this, based on the example of V1309 Sco we may expect changes in the average brightness and the light curve shape when very close to $t_0$.

A series of high precision spectroscopic observations of the system velocity of the binary can be used to confirm the existence of component C.
Based on the timing variation, we expect sinusoidal variation with a semiamplitude of 1.9~\kms\ on the 590~d period.
The same observations can also test the alternative hypothesis of a degenerate component D.
This hypothesis predicts a gradual increase in system velocity of 10~\kms\ over the next several year followed by a long period of no change.

A detailed study of the time variation of starspots would be a probe of the convective layer.
As dissipation of orbital energy increases with the approach of merger, one might expect a disruption in the present pattern of convection.
High quality spectroscopic measurements would be a valuable supplement to the photometric light curves as they can resolve the ambiguity of which star a spot is on.
Also, radio and X-ray emission might be expected from the corona surrounding the contact binary and could provide and independent probe of magnetic activity.

Reanalysis of the photometric and spectroscopic data with a more realistic physical model than used in this paper would improve estimates of physical properties and their uncertainties.
The soon-to-be-released second version of the Phoebe analysis package\footnote{URL: http:phoebe-project.org} (as described by \citet{2016arXiv160908135P}) will be well suited for this.

Finally, more observations of the orbital timing of other contact binaries are needed.
For example, if new timing measurement of the systems in the Kepler eclipsing binary catalog show the observation of an increasingly negative period derivative in KIC 9832227 is unique, we might conclude some new physical behavior is being exhibited independent of the merging star hypothesis.
Also, other precursor candidates should be sought in new, large surveys now coming online.
The key is to identify contact binaries promptly and follow up candidates with large period derivatives with intensive monitoring to be sure precursors are identified before they merge.

\acknowledgments

This paper makes use of data from the first public release of the WASP data (Butters et al. 2010) as provided by the WASP consortium and services at the NASA Exoplanet Archive, which is operated by the California Institute of Technology, under contract with the National Aeronautics and Space Administration under the Exoplanet Exploration Program.

Partial support was received from the Hubert A. Vander Plas Memorial Student Research Fellowship (CA and DMVN), the Dragt Family (EMC), the John Van Zytveld Student Summer Research Fellowship (BJ and CJS), NSF grant AST-1412845 (HAK), a Calvin Research Fellowship (LAM), a Calvin Integrated Science Research Institute grant (DMVN) and a Michigan Space Grant Consortium fellowship (DMVN).
We thank  Russet McMillan and Suzanne Hawley for generosity in obtaining needed observing time at APO.
We thank William Chick, Grace Olivier, Heather Wernke, Rebecca Sorber, Michael Lundquist, and Stephan Munari for obtaining data at WIRO in support of this program.
LAM would like to thank Jean McKeever for directing us to the broadening function and assisting in its computation.
JPS would like to thank Sergio Blanco-Cuaresma for his helpful guidance with {\it iSpec}.

\facilities{Kepler, ARC, WIRO, ASAS, Super-WASP}

%\bf next line
{\software{ACP, Astropy \citep{2013A&A...558A..33A}, Binary Maker 3.03 \citep{2002AAS...201.7502B}, IRAF, iSpec \citep{2014A&A...569A.111B}, MaxIm DL}

\bibliography{KIC9832227}

\clearpage
\begin{deluxetable}{lllrlrll}
\tabletypesize{\footnotesize}
\tablecolumns{8}
\tablewidth{0pt}
\tablecaption{Calvin College Observing Log \label{tabCalvinLog}}
\tablehead{\colhead{Epoch} & \colhead{UT first} & \colhead{UT last} & \colhead{Nights} & \colhead{Site} & \colhead{\# R/r} & \colhead{Filters} & \colhead{Complete}}
\startdata
    2013-01 & 2013 Jun 14 & 2013 Jun 19 & 3     & MI    & 144   & BR    & no \\
    2013-02 & 2013 Jul 11 & 2013 Jul 25 & 5     & MI    & 240   & BR    & yes \\
    2013-03 & 2013 Oct 7 & 2013 Oct 15 & 5     & NM    & 1047  & BR    & yes \\
    2014-01 & 2014 Mar 9 & 2014 Mar 19 & 6     & NM    & 804   & R     & yes \\
    2014-02 & 2014 Jun 7 & 2014 Jun 7 & 1     & MI    & 163   & R     & no \\
    2014-03 & 2014 Jun 29 & 2014 Jun 30 & 2     & NM    & 462   & R     & no \\
    2014-04 & 2014 Aug 3 & 2014 Aug 7 & 3     & MI, NM & 1052  & R     & no \\
    2014-05 & 2014 Aug 30 & 2014 Sep 1 & 3     & NM    & 1338  & R     & yes \\
    2014-06 & 2014 Sep 23 & 2014 Oct 6 & 8     & NM    & 1483  & R     & yes \\
    2014-07 & 2014 Oct 13 & 2014 Oct 15 & 3     & NM    & 172   & R     & yes \\
    2014-08 & 2014 Oct 25 & 2014 Oct 31 & 6     & NM    & 196   & griz  & yes \\
    2014-09 & 2014 Nov 4 & 2014 Nov 9 & 5     & NM    & 141   & griz  & yes \\
    2015-01 & 2015 Mar 28 & 2015 Apr 1 & 4     & NM    & 895   & R     & yes \\
    2015-02 & 2015 Apr 10 & 2015 Apr 15 & 5     & NM    & 756   & r     & yes \\
    2015-03 & 2015 May 1 & 2015 May 1 & 1     & NM    & 249   & r     & no \\
    2015-04 & 2015 May 11 & 2015 May 22 & 4     & NM    & 1566  & R     & yes \\
    2015-05 & 2015 Jun 2 & 2015 Jun 8 & 5     & NM    & 415   & VRI   & yes \\
    2015-06 & 2015 Jun 18 & 2015 Jun 21 & 3     & NM    & 244   & VRI   & yes \\
    2015-07 & 2015 Jul 28 & 2015 Jul 31 & 3     & MI    & 439   & R     & no \\
    2015-08 & 2015 Aug 19 & 2015 Aug 23 & 3     & MI, NM & 630   & VRI   & yes \\
    2015-09 & 2015 Sep 9 & 2015 Sep 12 & 3     & NM    & 263   & VRI   & yes \\
    2015-10 & 2015 Sep 25 & 2015 Sep 28 & 3     & NM    & 1110  & R     & yes \\
    2015-11 & 2015 Nov 14 & 2015 Nov 24 & 5     & NM    & 1121  & R     & yes \\
    2016-01 & 2016 Jan 29 & 2016 Feb 8 & 5     & NM    & 434   & R     & no \\
    2016-02 & 2016 Feb 9 & 2016 Feb 15 & 7     & NM    & 731   & R     & yes \\
    2016-03 & 2016 Feb 16 & 2016 Feb 22 & 6     & NM    & 667   & R     & no \\
    2016-04 & 2016 Feb 23 & 2016 Mar 4 & 10    & NM    & 1505  & R     & yes \\
    2016-05 & 2016 Mar 9 & 2016 Mar 19 & 4     & NM    & 822   & R     & no \\
    2016-06 & 2016 Mar 27 & 2016 Apr 5 & 2     & NM    & 521   & R     & no \\
    2016-07 & 2016 Apr 19 & 2016 Apr 26 & 6     & NM    & 1224  & R     & no \\
    2016-08 & 2016 May 12 & 2016 May 16 & 3     & NM    & 940   & R     & yes \\
    2016-09 & 2016 May 21 & 2016 May 28 & 6     & NM    & 1296  & R     & yes \\
    2016-10 & 2016 Jun 3 & 2016 Jun 8 & 3     & NM    & 108   & R     & no \\
    2016-11 & 2016 Jun 14 & 2016 Jun 20 & 7     & NM    & 1692  & R     & yes \\
    2016-12 & 2016 Jul 4 & 2016 Jul 10 & 6     & NM    & 1034  & R     & yes \\
    2016-13 & 2016 Jul 13 & 2016 Jul 17 & 3     & NM    & 812   & R     & no \\
    2016-14 & 2016 Aug 7 & 2016 Aug 10 & 4     & MI    & 791   & R     & yes \\
    2016-15 & 2016 Sep 2 & 2016 Sep 5 & 4     & MI    & 782   & R     & yes \\
\enddata
\end{deluxetable}

\clearpage
\begin{deluxetable}{ccccc}
\tabletypesize{\normalsize}
\tablecolumns{5}
\tablewidth{0pt}
\tablecaption{KIC 9832227 Orbital Period \label{tabKIC-P}}
\tablehead{\colhead{MJD} & \colhead{Period} & \colhead{Sigma} & \colhead{N} & \colhead{Data set} \\
  & \colhead{(d)} & \colhead{(d)} & & }
\startdata
    52981 & 0.45795348 & 0.00000018 & 5     & Pre-Kepler \\
    55689 & 0.45794852 & 0.00000013 & 71    & Kepler \\
    57061 & 0.45794121 & 0.00000034 & 25    & Calvin \\
\enddata
\end{deluxetable}

\clearpage
\begin{deluxetable}{lccc}
\tabletypesize{\scriptsize}
\tablecolumns{4}
\tablewidth{0pt}
\tablecaption{Summary of KIC 9832227 Spectroscopic Observations \label{tabspecobs}}
\tablehead{\colhead{UT Date} & \colhead{Int. Time} & \colhead{\# Spectra} & \colhead{Observatory} \\
  & \colhead{(s)} & & }
\startdata
2015 May 31 & 300 & 1 & WIRO$^{\rm{a}}$ \\
2015 Jun 02 & 300 & 4 & WIRO \\
2015 Jun 03 & 600 & 1 & WIRO \\
2015 Jun 03 & 300 & 2 & WIRO \\
2015 Jun 04 & 300 & 1 & WIRO \\
2015 Jun 08 & 600 & 6 & WIRO \\
2015 Jun 09 & 600 & 12 & WIRO \\
2015 Jun 13 & 600 & 15 & WIRO \\
2015 Jun 20 & 600 & 7 & WIRO \\
\hline
2012 Jul 18 & 1200 & 2 & APO$^{\rm{b}}$ \\
2012 Jul 21 & 1200 & 1 & APO \\
2012 Oct 04 & 1200 & 2 & APO \\
2012 Oct 24 & 600 & 2 & APO \\
2012 Oct 25 & 600 & 2 & APO \\
2012 Oct 28 & 1200 & 1 & APO \\
2013 Mar 25 & 1200 & 1 & APO \\
2015 Mar 28 & 1800 & 5 & APO \\
2015 Apr 15 & 1800 & 2 & APO \\
2015 May 31 & 900 & 8 & APO \\
2015 Jun 18 & 1800 & 1 & APO \\
2015 Jun 18 & 900 & 5 & APO \\
2015 Jun 21 & 900 & 18 & APO \\
\enddata
\tablecomments{$^{\rm{a}}$Wyoming Infrared Observatory; $^{\rm{b}}$Apache Point Observatory}
\end{deluxetable}

\clearpage
\begin{deluxetable}{lrrr}
\tabletypesize{\footnotesize}
\tablecolumns{4}
\tablewidth{0pt}
\tablecaption{KIC 9832227 Binary Physical Parameters \label{tabPhysicalParameters}}
\tablehead{\multicolumn{4}{c}{Stellar Parameters}}
\startdata
% & \multicolumn{2}{c}{Stellar Parameters} & \\
%\hline
 & Primary & Secondary & Total \\
    $m/M_\odot$  & 1.395(11) & 0.318(5) & 1.714(12) \\
    $L/L_\odot$  & 2.609(55) & 0.789(17) & 3.398(72) \\
    $R_V/R_\odot$  & 1.581(9) & 0.830(5) &  \\
    $T$ (K) & 5800(59)  & 5920(63)  & 5828(58) \\
    $v\sin i$ (\kms ) & 149.7(8) & 84.7(5)  &  \\
    $\log g$ & 4.19(1)  & 4.10(1)  &  \\
\hline
 \multicolumn{4}{c}{System Parameters} \\
\hline
    $i$ (deg) & 53.19(10) &       &  \\
    $f$ & 0.430(15) &       &  \\
    $T_B$/$T_A$ & 1.021(6) & & \\
    mass ratio & 0.228(3) &       &  \\
    area ratio & 0.279(6) &       &  \\
    light ratio ($R$) & 0.307(10) &       &  \\
    distance (pc) & 565(16) &       &  \\
    E(B--V) (mag) & 0.03(2) &  &  \\
    $a$ ($R_\odot$) & 2.992(15) &       &  \\
    $P_0$ (days) & 0.4579615(9) &       &  \\
    $\tau$ (days) & 0.114(14) &       &  \\
    $t_0$ (year) & 2022.2(7) &       &  \\
    BJD$_0$ & MJD 55688.49913(2) &       &  \\
\hline
 \multicolumn{4}{c}{Coordinates} \\
\hline
    RA, Dec (J2000) & 19$^{\rm h}$29$^{\rm m}$15\fs 954 & +46\degr 37\arcmin 19\farcs 88 & \\
    Galactic (\emph{l}, \emph{b}) & 78\fdg 8007 & 13\fdg 3557 & \\
\enddata
\end{deluxetable}

\clearpage
\begin{figure}
\begin{center}
\plotone{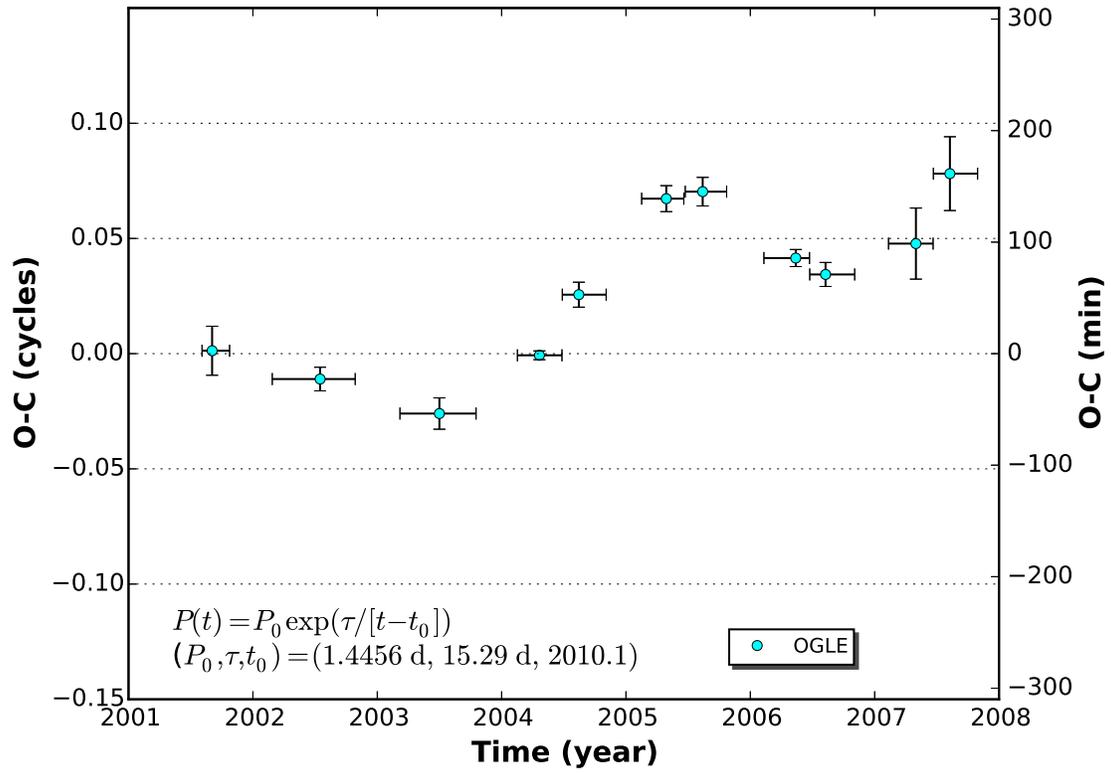}
\caption{The residual of the OGLE orbital phase of V1309 Sco relative to the exponential ephemeris of Te11.  The zero point is chosen to be the first datum in 2004.}\label{figV1309Sco-OMCresid}
\end{center}
\end{figure}

\clearpage
\begin{figure}
\begin{center}
\plotone{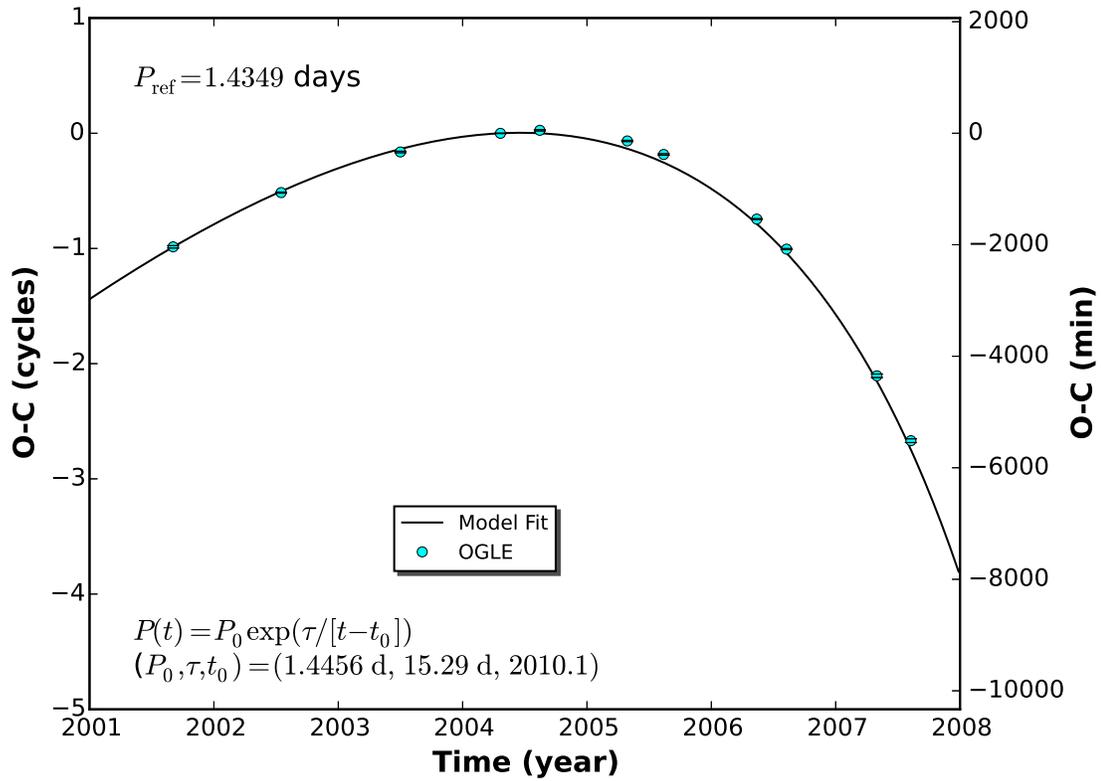}
\caption{The OGLE orbital phase of V1309 Sco (circles) minus that computed from a linear reference ephemeris. Uncertainties are smaller than the symbols. The exponential formula is indicated by a black line.}\label{figV1309Sco-OMC}
\end{center}
\end{figure}

\clearpage
\begin{figure}
\begin{center}
\plotone{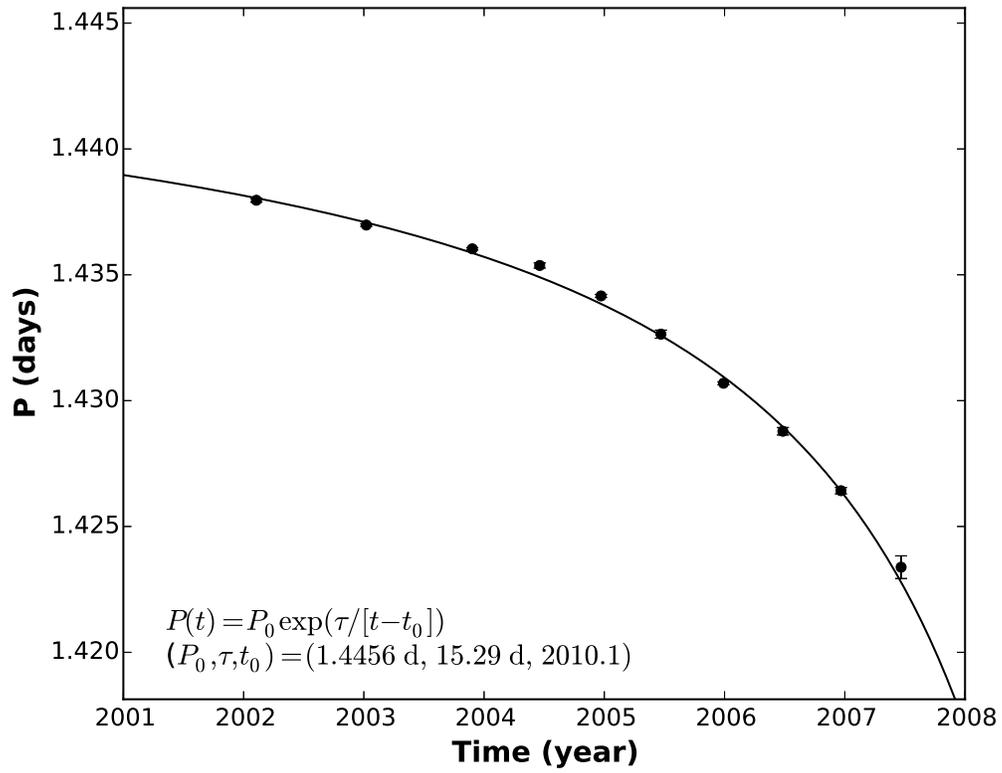}
\caption{The OGLE orbital period of V1309 Sco (circles) versus time. The exponential formula is indicated by a black line.  The period is plotted with $P_0$, the asymptotic value, as maximum.}\label{figV1309Sco-P}
\end{center}
\end{figure}

\clearpage
\begin{figure}
\begin{center}
\plotone{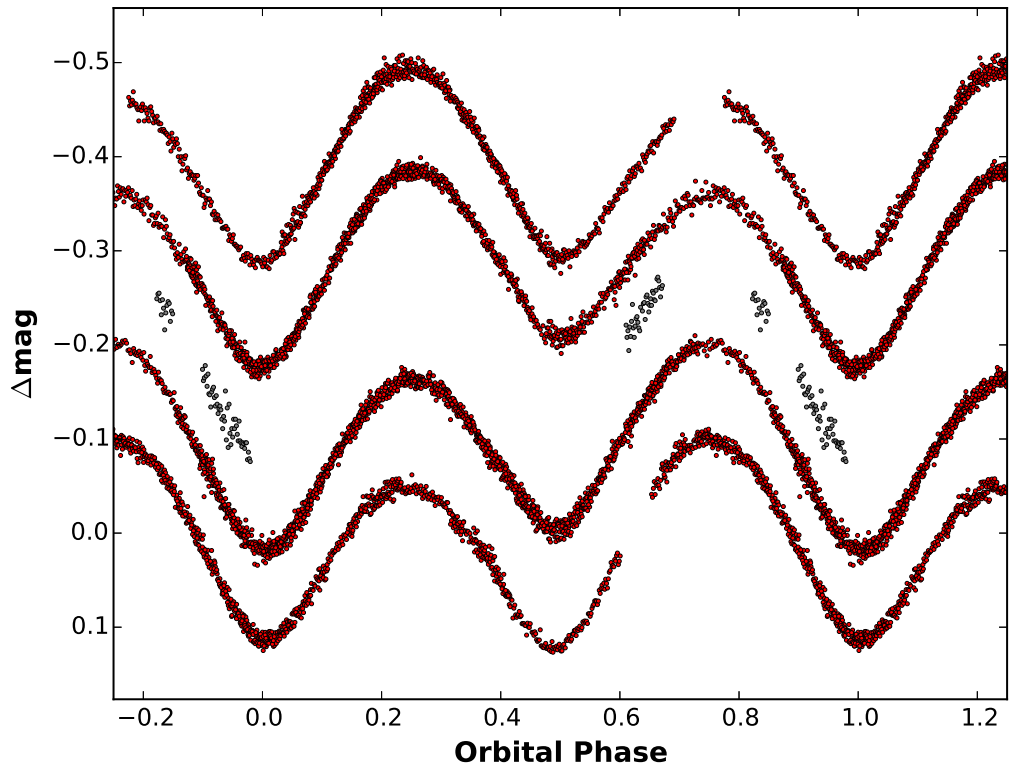}
\caption{A sample of Calvin College photometry of KIC 9832227 versus orbital phase for five epochs of observations (2016-08 to 2016-12, which span 2016 May to July, see Table \ref{tabCalvinLog}).  The earliest epoch is at the top with subsequent ones offset down by 0.1 mag increments for visibility.  Red symbols indicate epochs with complete phase coverage; grey symbols indicate incomplete coverage.}\label{figCalvinLCs}
\end{center}
\end{figure}

\clearpage
\begin{figure}
\begin{center}
\plotone{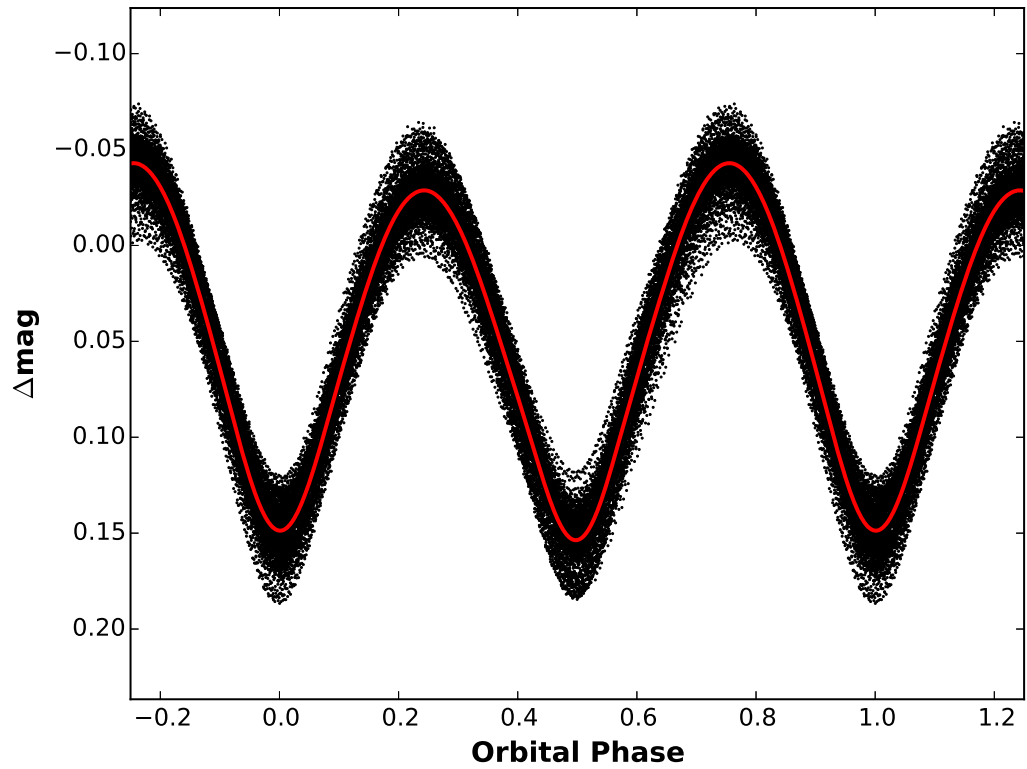}
%\bf spanning ... January
\caption{Kepler spacecraft photometry of KIC 9832227 spanning 2009 May to 2013 January versus orbital phase (black dots).  The time averaged light curve (as fit by a 10th order Fourier series) is given as as red line.  The spread of the data about the line is due to time variability of the light curve shape.}\label{figKeplerLC}
\end{center}
\end{figure}

\clearpage
\begin{figure}
\begin{center}
\plotone{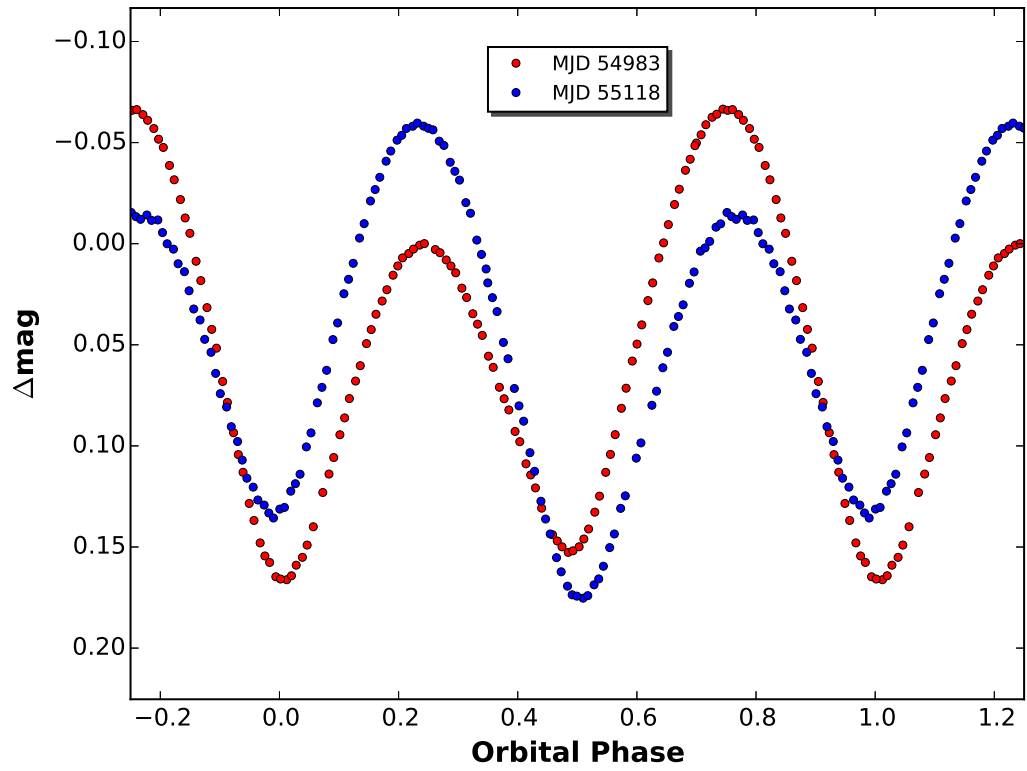}
\caption{Kepler spacecraft photometry of KIC 9832227 versus orbital phase for five orbits beginning on MJD 54983 (red circles) and five orbits beginning on MJD 55118 (blue circles).  The internal consistency of the two sets illustrates the high quality of the Kepler data, while the large difference between the sets illustrates the range of variation of light curve shape.}\label{figSampleLCs}
\end{center}
\end{figure}

\clearpage
\begin{figure}
\begin{center}
\plotone{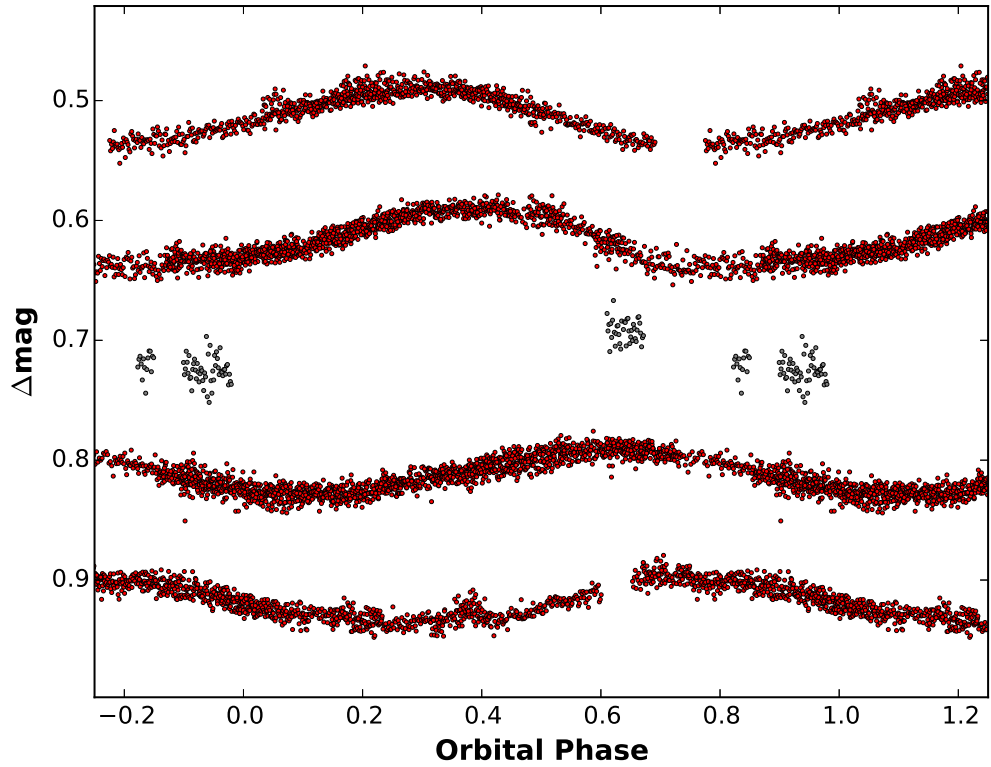}
\caption{A sample of residuals of Calvin College photometry relative to the average Kepler light curve of KIC 9832227 (Figure \ref{figKeplerLC}) versus orbital phase for five epochs of observations (2016-08 to 2016-12, which span 2016 May to July, see Table \ref{tabCalvinLog}).  The earliest epoch is at the top with subsequent ones offset down by 0.1 mag increments for visibility.  Red symbols indicate epochs with complete phase coverage; grey symbols indicate incomplete coverage. The residual shapes in these epochs are approximately sinusoidal, indicating the dominance of a single starspot region.  The phase of the sinusoid drifts gradually, indicating migration of the longitude of the starspots.}\label{figCalvinResids}
\end{center}
\end{figure}

\clearpage
\begin{figure}
\begin{center}
\plottwo{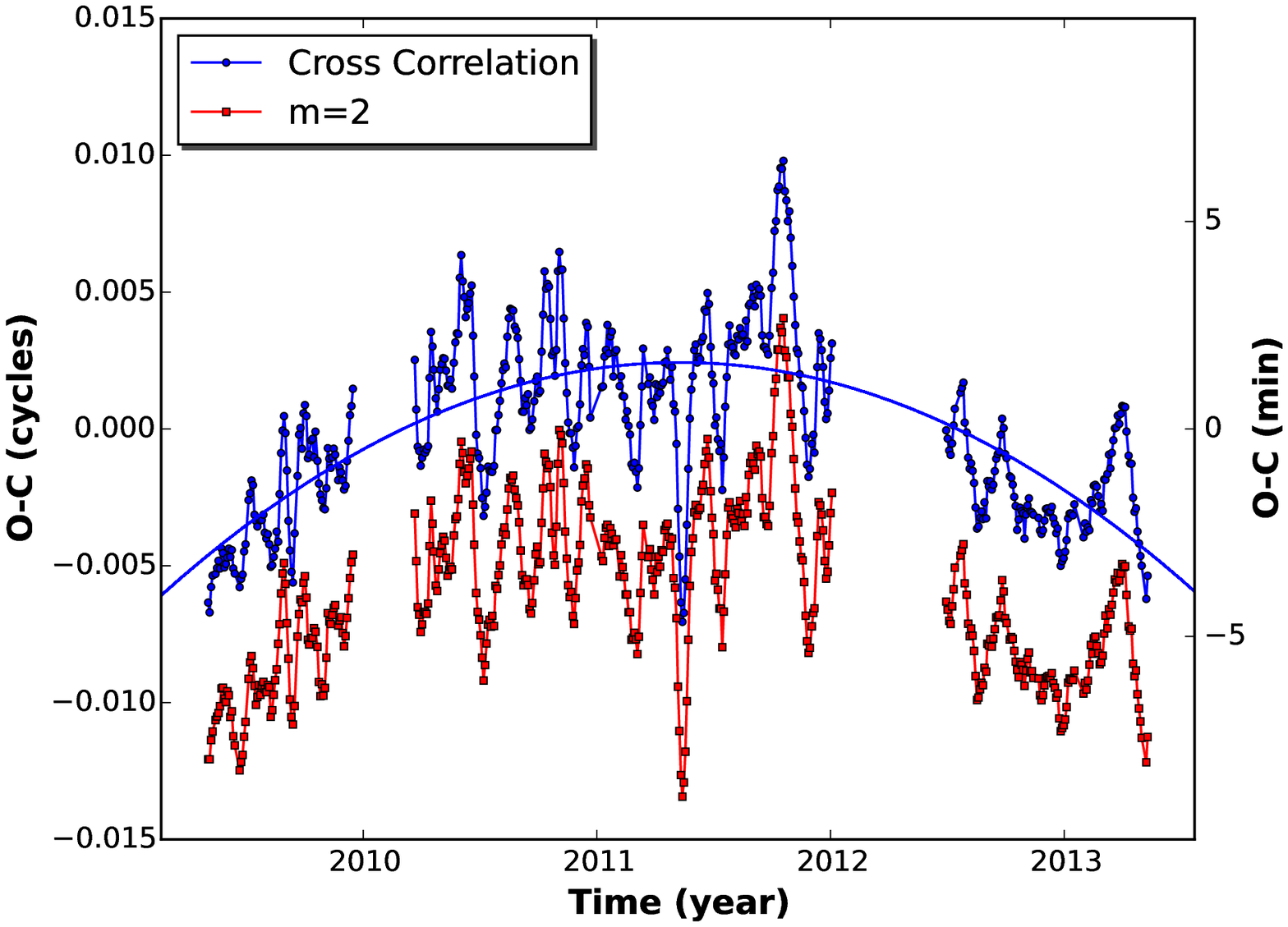}{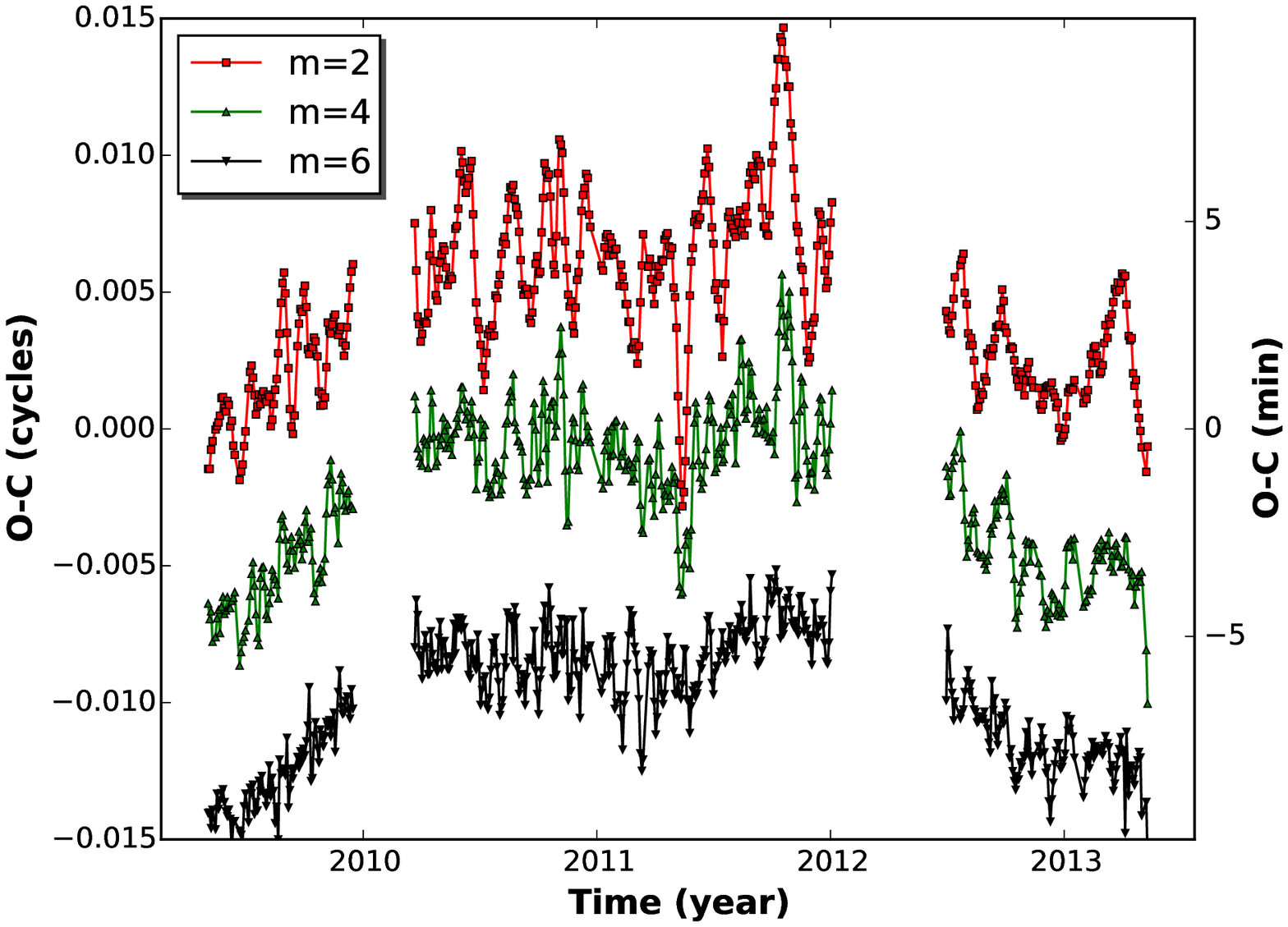}
\caption{(a) The Kepler orbital phase of KIC 9832227 relative to the linear ephemeris of \citet{2014AJ....147...45C} as computed with a cross correlation (blue symbols) and from the phase of the second Fourier order.  The blue parabola is a constant period derivative fit to the cross correlation data. (b) The same computed from the phases of the second, fourth, and sixth Fourier orders (top to bottom), with vertical offsets inserted for clarity.}\label{figConroy}
\end{center}
\end{figure}

\clearpage
\begin{figure}
\begin{center}
\plottwo{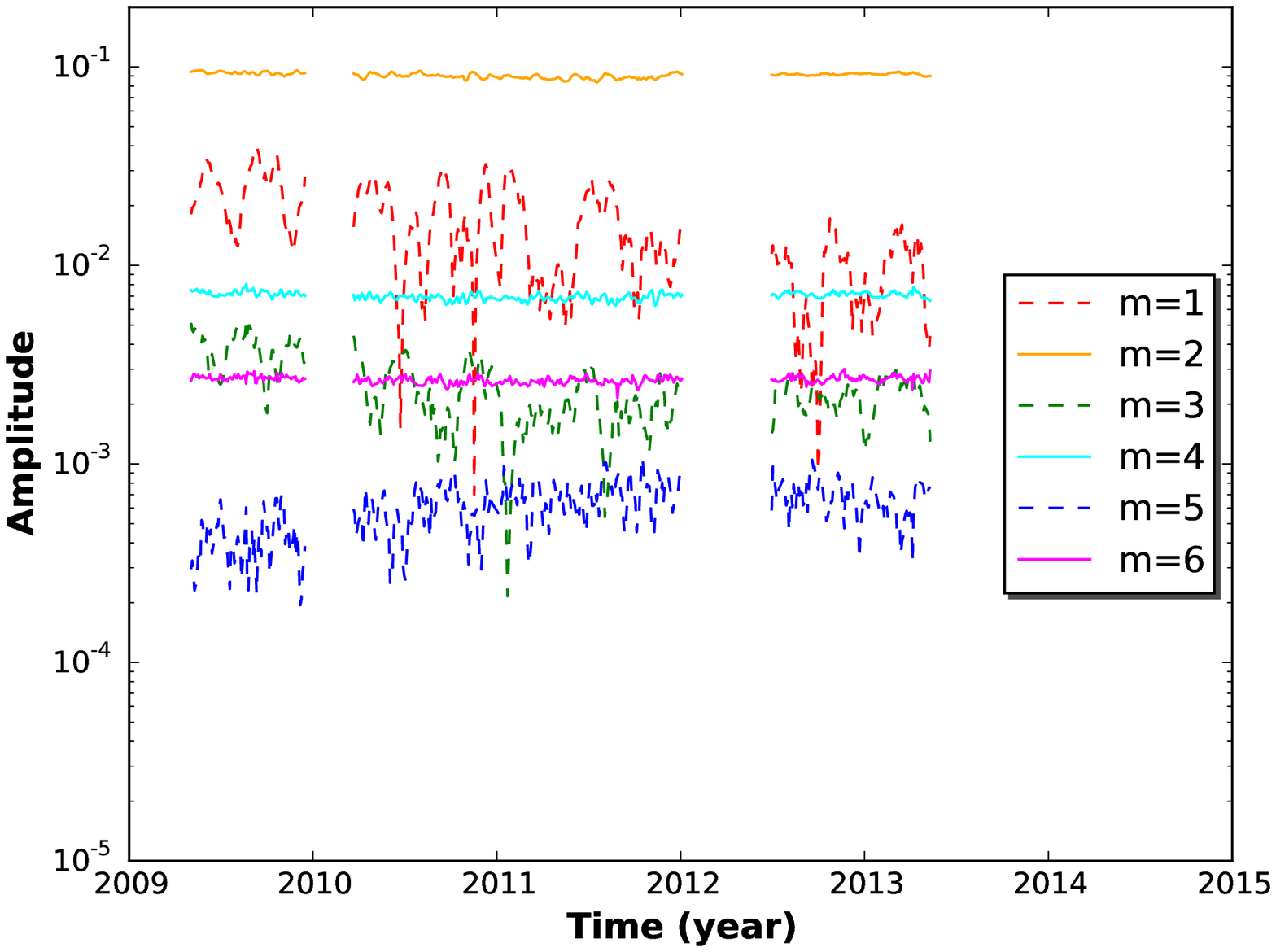}{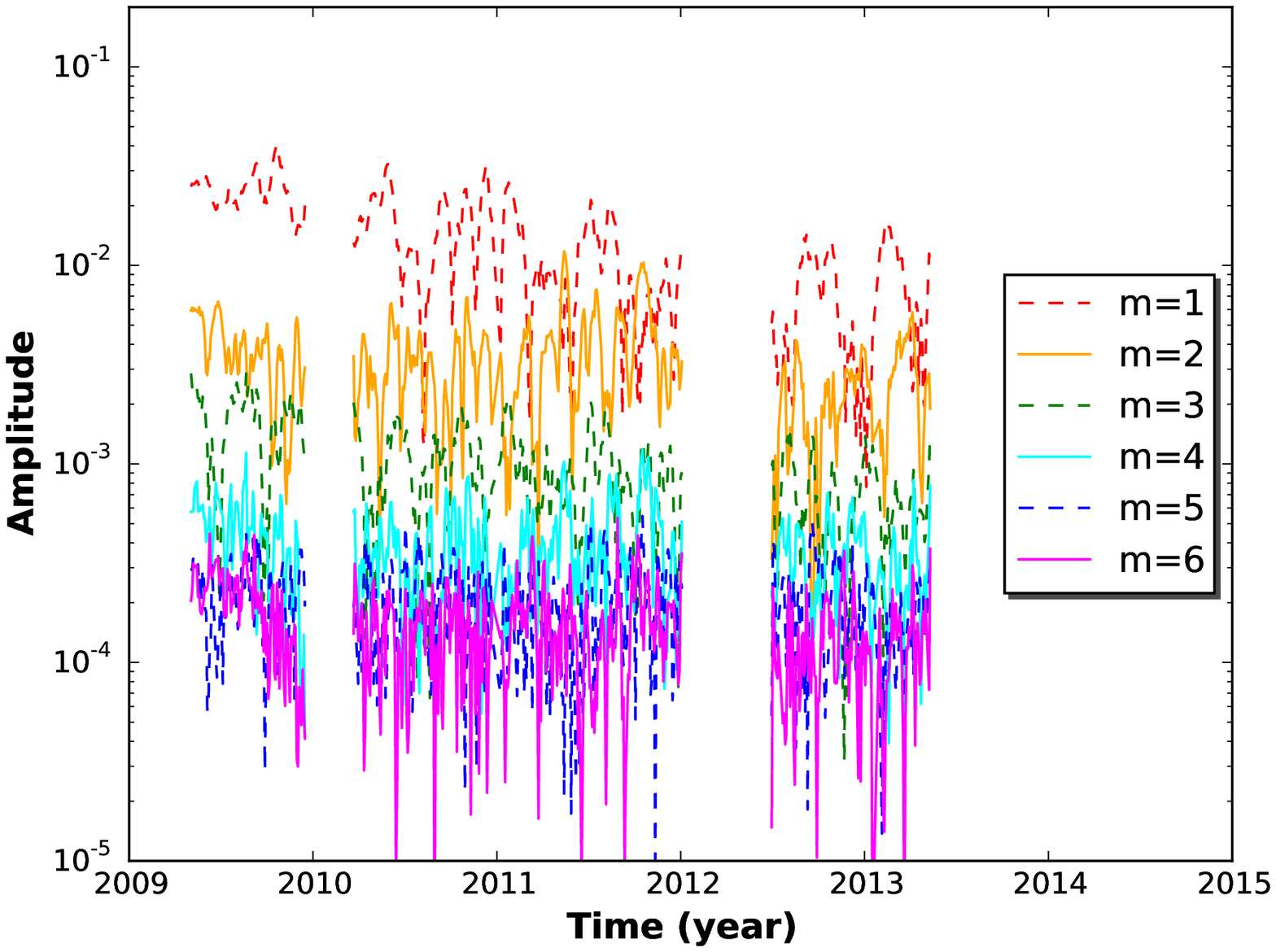}
\caption{(a) Amplitudes of the first six Fourier orders computed for five orbit time segments of the Kepler data for KIC 9832227, with colors indicated in the legend. (b) The same for Kepler data residuals relative to the time averaged light curve.}\label{figFALC}
\end{center}
\end{figure}

\clearpage
\begin{figure}
\begin{center}
\plotone{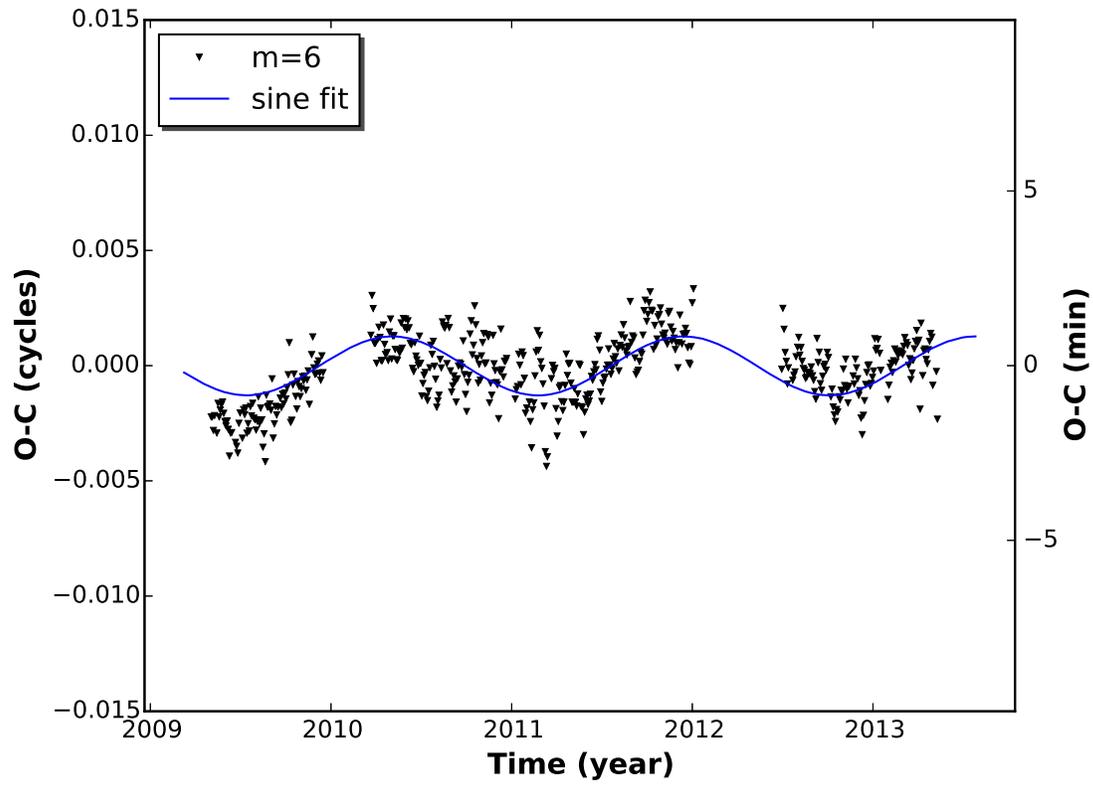}
\caption{The Kepler orbital phase of KIC 9832227 relative to the exponential ephemeris of \S \ref{subsecExp} computed with sixth Fourier order phases.  A sinusoidal fit to the data is plotted as a blue line.}\label{figSmallBodyOMC}
\end{center}
\end{figure}

\clearpage
\begin{figure}
\begin{center}
\plotone{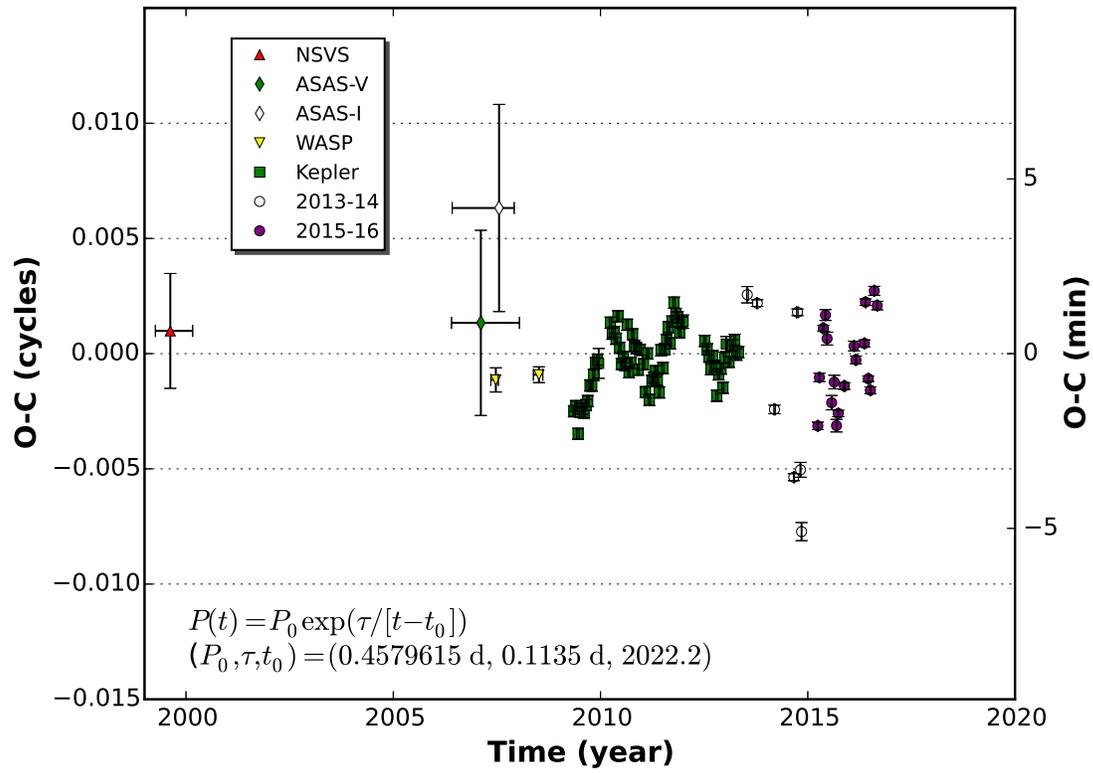}
\caption{The residual of the orbital phase of KIC 9832227 relative to the exponential ephemeris of \S \ref{subsecExp}.  The zero point is chosen to be the average of the sine wave from Figure \ref{figSmallBodyOMC}.}\label{figKIC-OMCresid}
\end{center}
\end{figure}

\clearpage
\begin{figure}
\begin{center}
\plotone{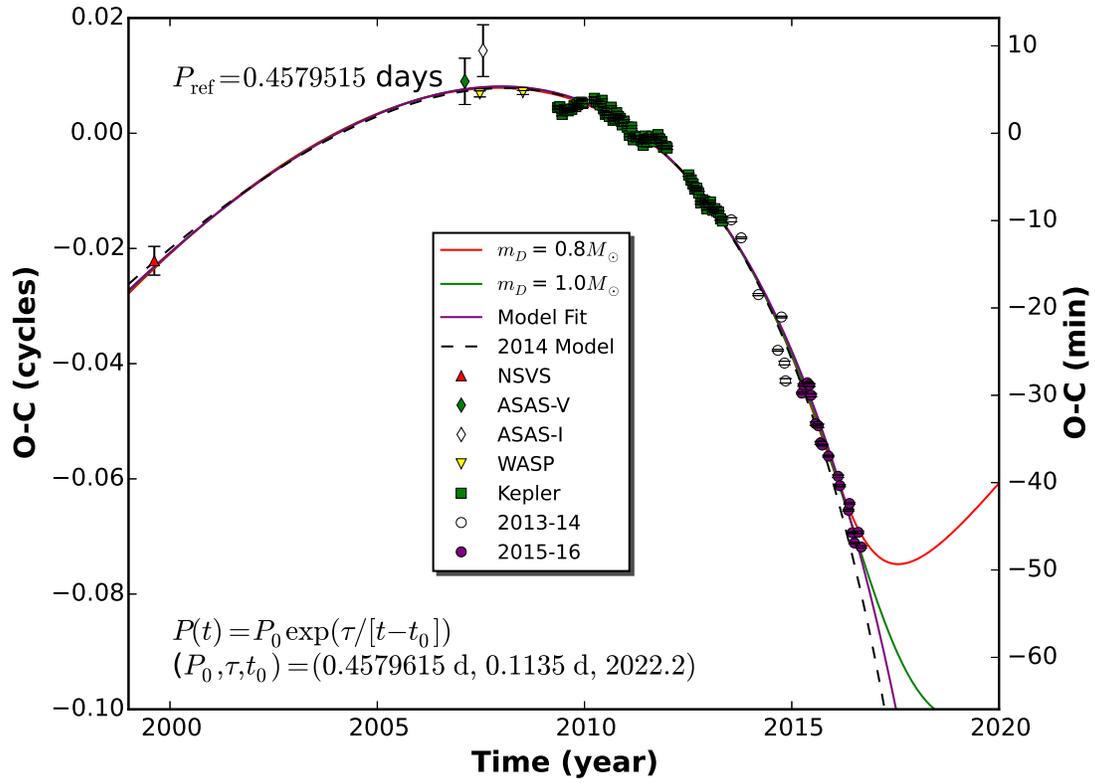}
\caption{The observed orbital phase of KIC 9832227 minus that computed from a linear reference ephemeris.  The exponential fit of Me15 is indicated by a dashed black line.
Data taken subsequently, filled purple circles, follow this line within its uncertainty.
The updated fit of \S \ref{subsecExp} is indicated by a solid purple line (with parameters given in the plot).
Alternative fits, based on light travel time delay due to the orbit of the triple system around a hypothetical component D, are shown as red and green lines (for assumed masses $m_D$ = 0.8 and 1.0 $M_\odot$, respectively) as described in  \S \ref{subsecCompD}.
The model parameters, ($m_D$,$P$,$e$,$t_p$,$\omega$), are (0.8~$M_\odot$,25.1~y,0.609,43 \degr, 2016.5) and (1.0~$M_\odot$,30.7~y,0.569,53 \degr, 2017.2), respectively.}\label{figKIC-OMC}
\end{center}
\end{figure}

\clearpage
\begin{figure}
\begin{center}
\plotone{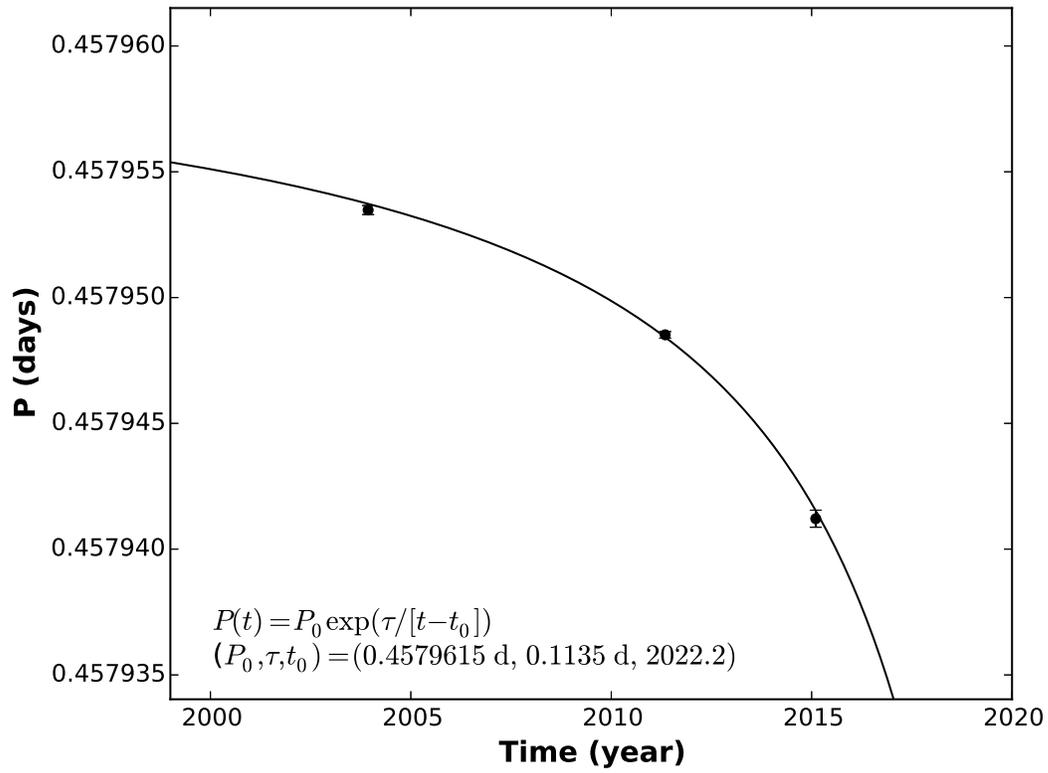}
\caption{Orbital period of KIC9832227 with time.
Data points are derived from pre-Kepler data, Kepler data, and Calvin College data, while the solid line is the exponential fit of the timing data in Figure \ref{figKIC-OMC} as described in \S \ref{subsecExp}.}\label{figKIC-P}
\end{center}
\end{figure}

\clearpage
\begin{figure}
\begin{center}
\plotone{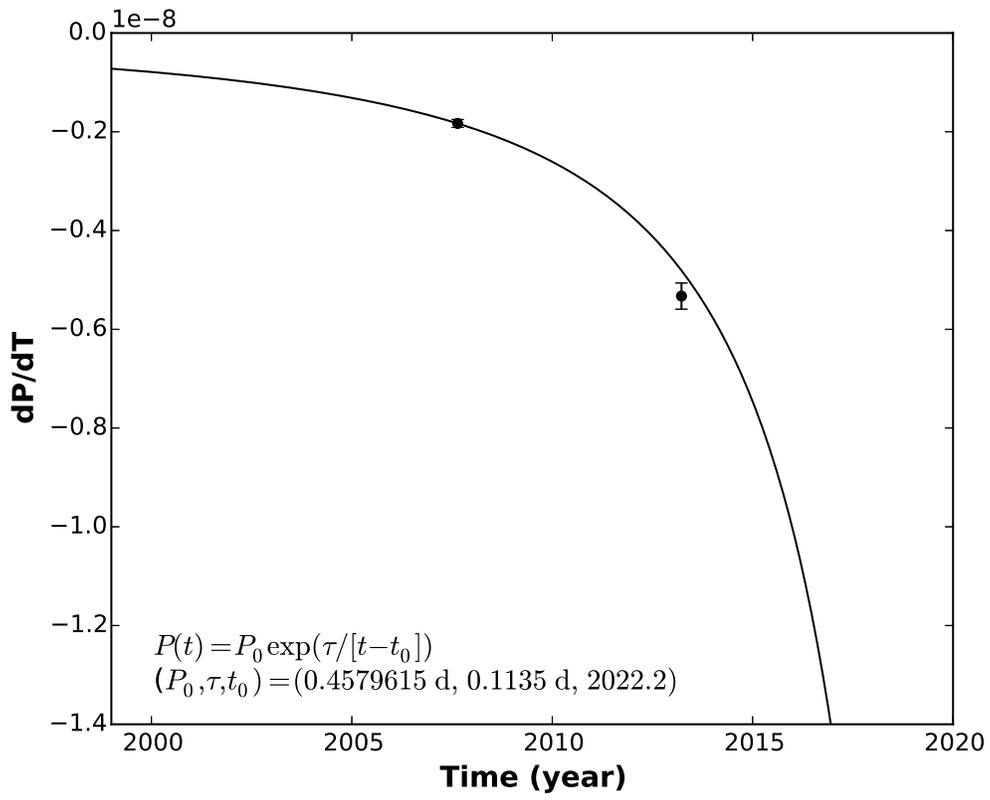}
\caption{Orbital period derivative of KIC 9832227 with time.
Data points are based on the differences in the periods in Figure \ref{figKIC-P}, while the solid line is the exponential fit of the timing data in Figure \ref{figKIC-OMC} as described in \S \ref{subsecExp}.}\label{figKIC-Pdot}
\end{center}
\end{figure}

\clearpage
\begin{figure}
\begin{center}
\plotone{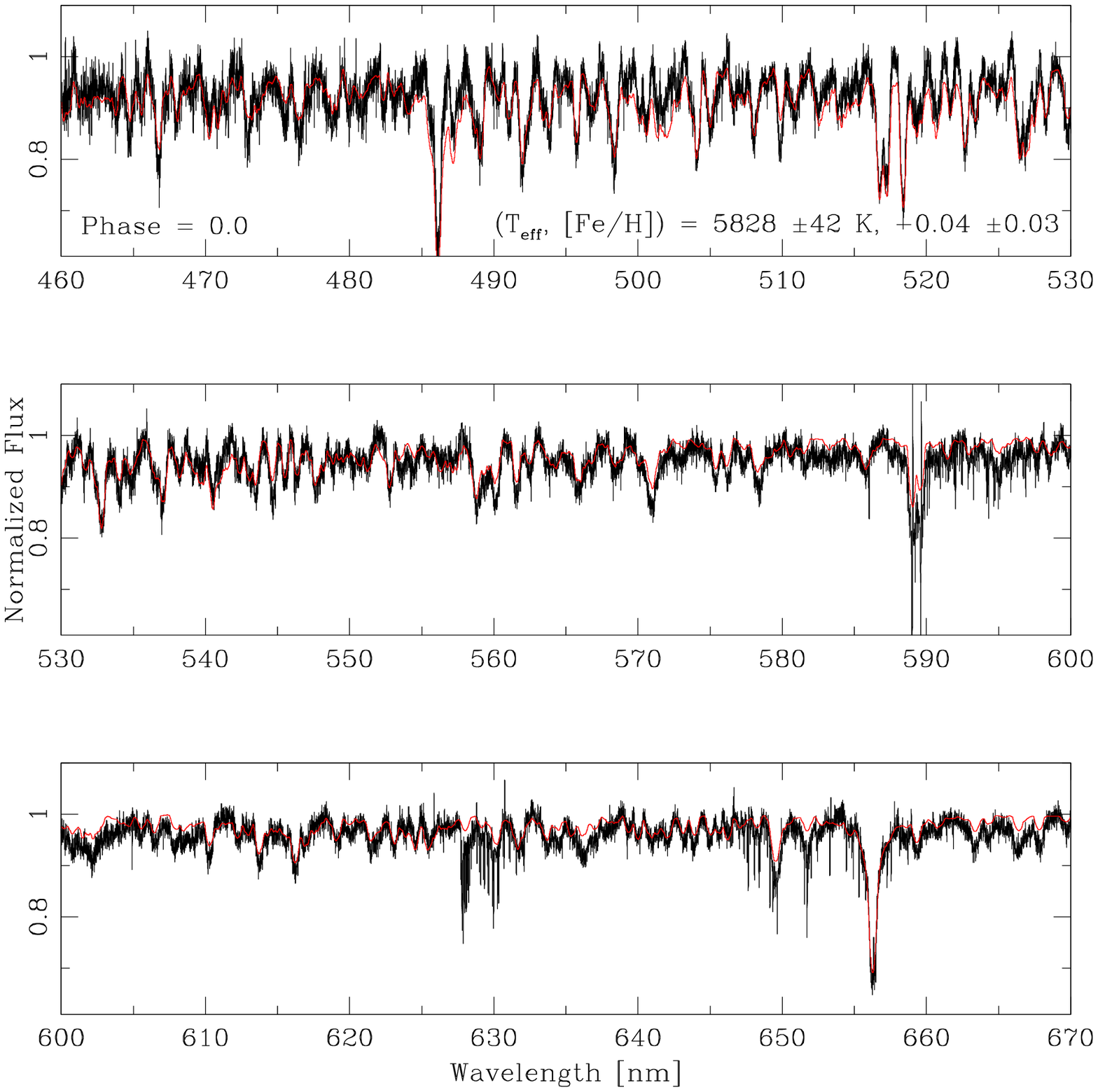}
\caption{Co-added APO spectrum at orbital phase $\phi$ = 0.0 (black), with the best-fit synthetic spectrum corresponding to $T_{\rm eff}$ = 5828 K and [Fe/H] = -0.04 shown in red.}\label{figspecapo00}
\end{center}
\end{figure}

\clearpage
\begin{figure}
\begin{center}
\plotone{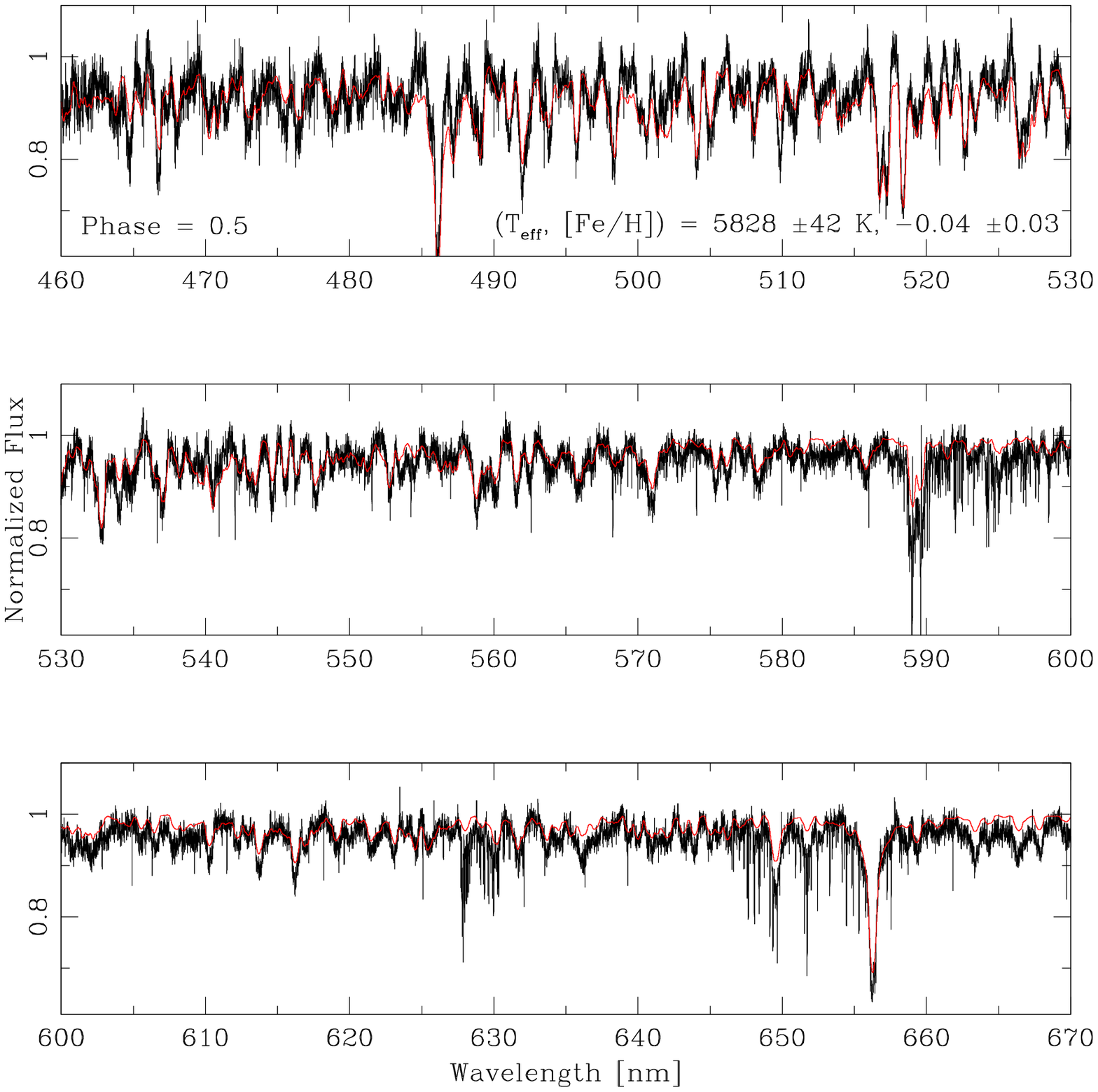}
\caption{Co-added APO spectrum at orbital phase $\phi$ = 0.5 (black), with the best-fit synthetic spectrum corresponding to $T_{\rm eff}$ = 5828 K and [Fe/H] = $-$0.04 shown in red.}\label{figspecapo05}
\end{center}
\end{figure}

\clearpage
\begin{figure}
\begin{center}
\plotone{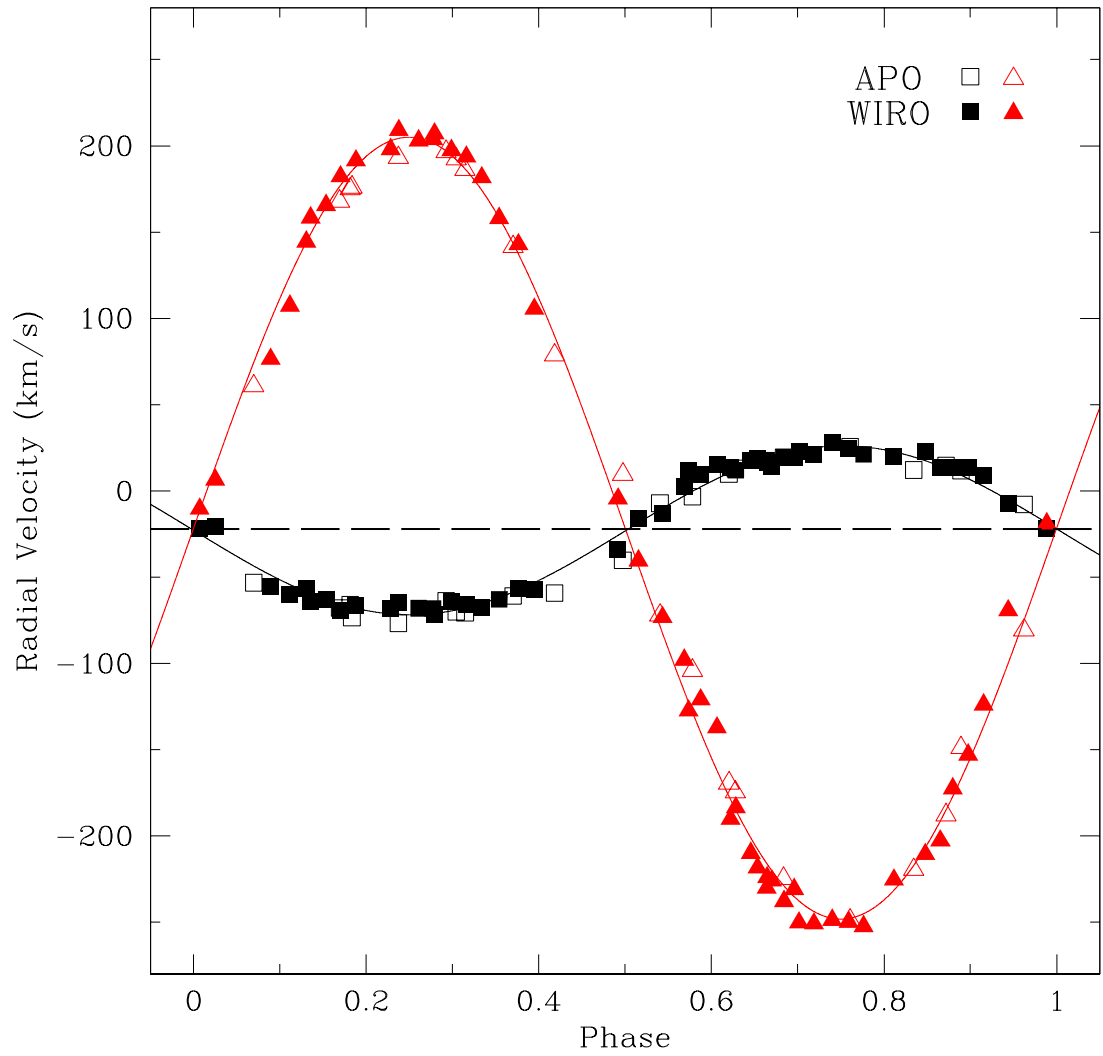}
%\bf Squares ... WIRO data
\caption{Radial velocity curves for KIC 9832227 components A and B derived from Gaussian fits to each component in the broadening function calculations.  Squares denote component A, triangles component B; open symbols denote APO data, filled symbols WIRO data.
Error bars for the data points have been omitted for clarity, but are typically 1--4~\kms \ and smaller than the size of the data points.}\label{figbfrvcurvegaussian}
\end{center}
\end{figure}

\clearpage
\begin{figure}
\begin{center}
\plottwo{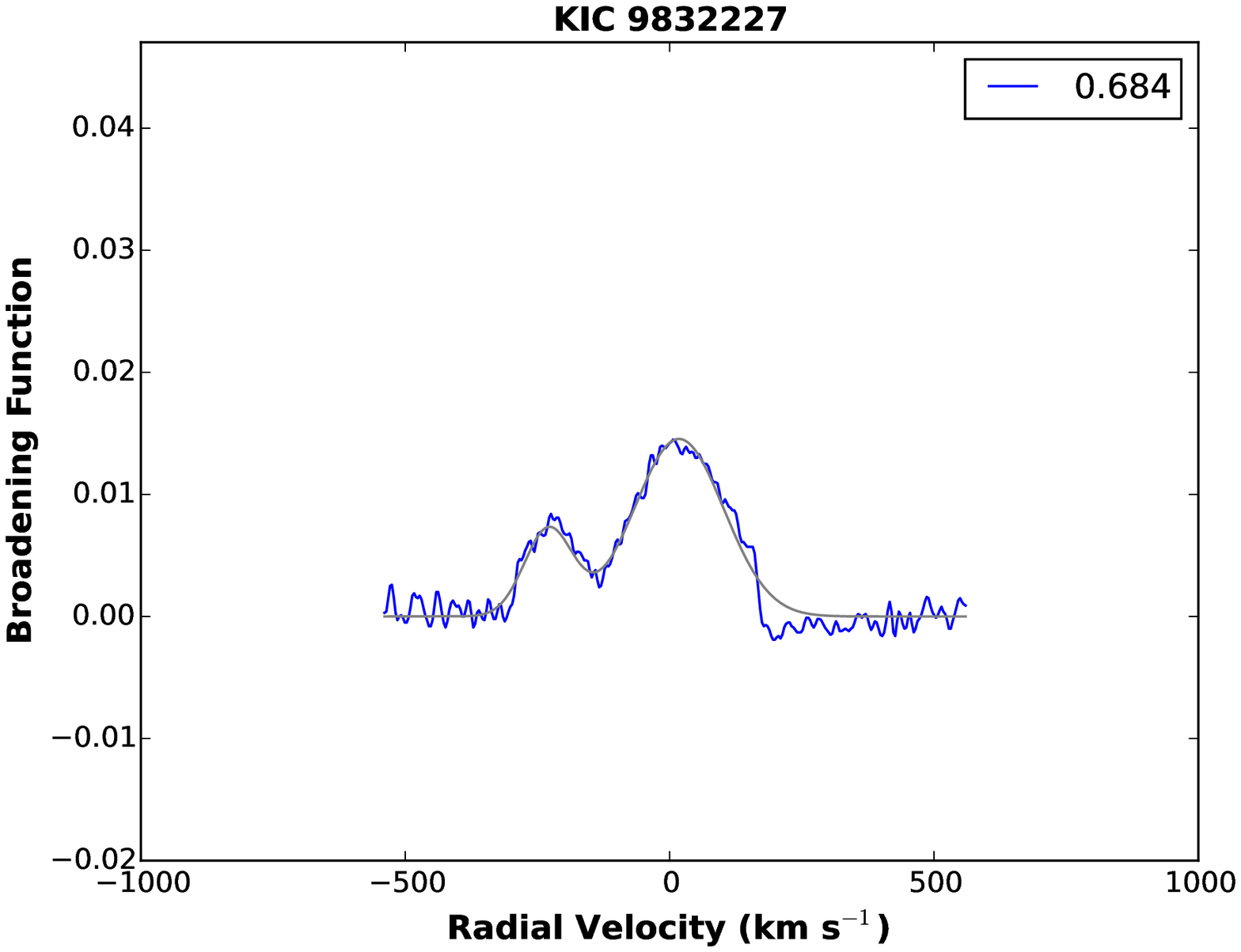}{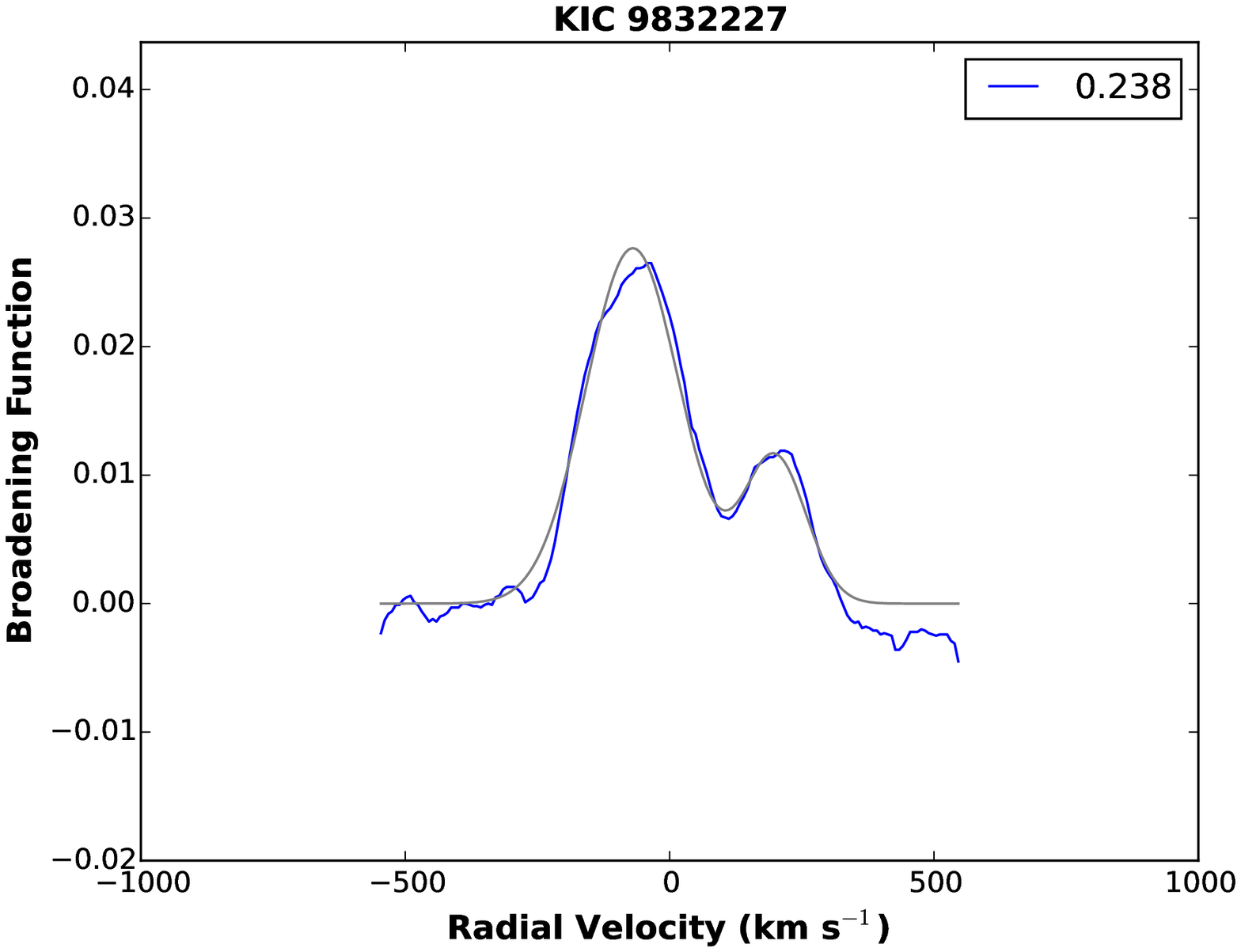}
\caption{Two illustrative broadening functions (blue) and model fits (gray): a) APO data at orbital phase 0.684. (b) WIRO data at orbital phase 0.238.}\label{figBFFit}
\end{center}
\end{figure}

\clearpage
\begin{figure}
\begin{center}
\plottwo{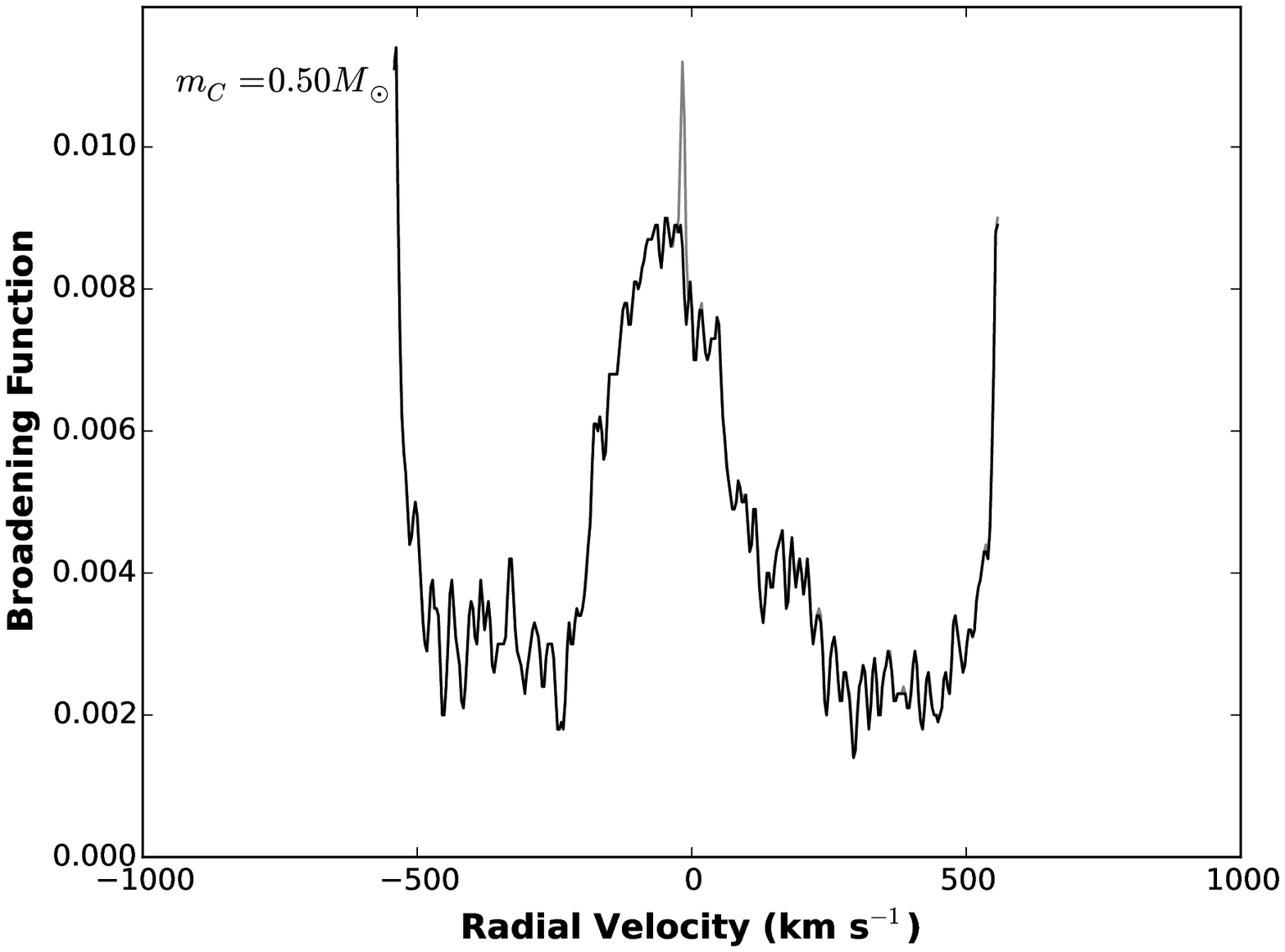}{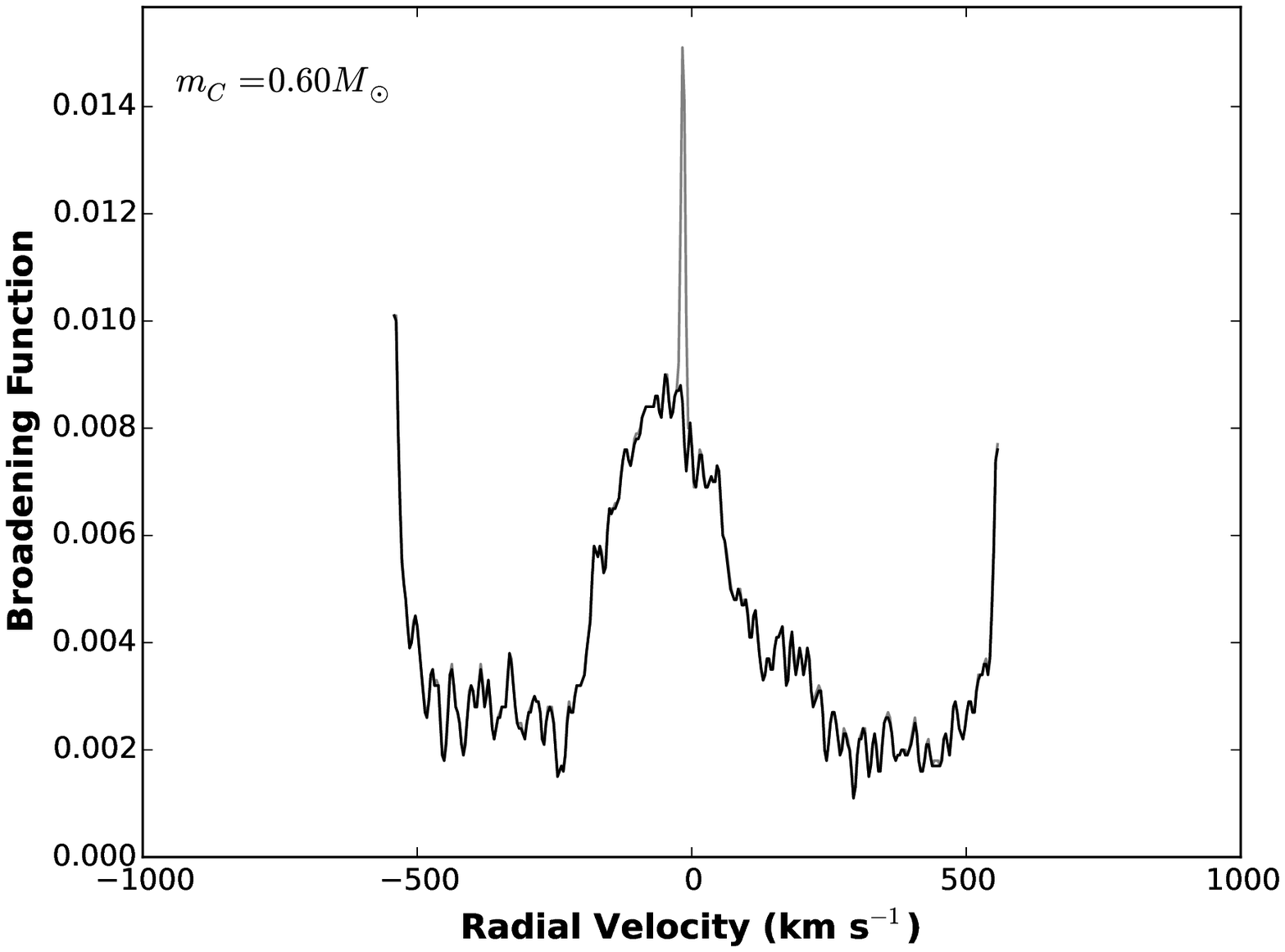}
\caption{a) Broadening function (black) of APO data from 2015 June 21 using a template for a 0.5 $M_\odot$ main sequence star (4000 K).  For comparison, the gray line is the same but with a scaled template spectrum added to the APO spectrum first. (b) same as (a) for an 0.6 $M_\odot$ star (4250 K).}\label{figThirdBodyBF}
\end{center}
\end{figure}

\clearpage
\begin{figure}
\begin{center}
\plotone{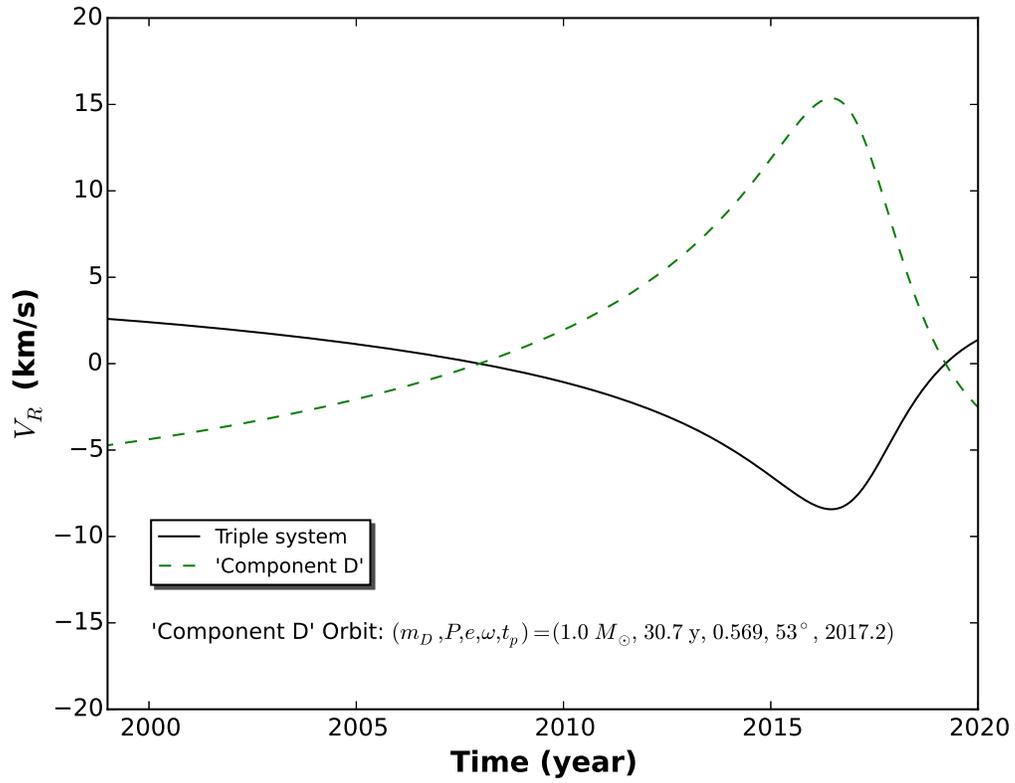}
\caption{The radial velocities associated with the alternative interpretation of the long term timing trend, light travel time delays due to the orbit of the triple system about a hypothesized component D (described in  \S \ref{subsecCompD}).
Velocity of the triple is marked with a solid black line, while that of component D is marked with a dashed green line.
The orbital parameters, which are labelled in the Figure, match the case marked with a green line in Figure \ref{figKIC-OMC}.}\label{figKIC-RVD}
\end{center}
\end{figure}

\end{document}